\documentclass[twocolumn,aps,pra,showpacs,raggedbottom,nobalancelastpage,amssymb,superscriptaddress]{revtex4-1}

\usepackage{graphicx}
\usepackage{amsmath}
\usepackage{amssymb}
\usepackage{bm}
\usepackage[usenames]{color}
\usepackage{amsfonts}
\usepackage{appendix}
\usepackage[breaklinks=true,colorlinks=true,linkcolor=blue,urlcolor=blue,citecolor=blue]{hyperref}

\usepackage{dsfont}
\graphicspath{{Figures/}}

\def\be{\begin{equation}}
	\def\ee{\end{equation}}

\begin{document}
	\title{Theory of hydrodynamic phenomena in optical mesh lattices}

	\author{Hannah M. Price}
	\address{School of Physics and Astronomy, University of Birmingham,
		Edgbaston Park Road, B15 2TT, West Midlands, United Kingdom}
	
	\author{Martin Wimmer}
	\author{Monika Monika}
	\address{Institute of Condensed Matter Theory and Optics Friedrich-Schiller-University Jena, Max-Wien-Platz 1,
		D-07743 Jena, Germany}
	
	\author{Ulf Peschel}
	\address{Institute of Condensed Matter Theory and Optics Friedrich-Schiller-University Jena, Max-Wien-Platz 1,
		D-07743 Jena, Germany}
	
	\author{Iacopo Carusotto}
	\address{INO-CNR BEC Center and Dipartimento di Fisica, Universit\`{a} di Trento, I-38123 Povo, Italy}

	\date{\today}
	\begin{abstract}
		Signatures of superfluid-like behaviour have recently been observed experimentally in a nonlinear optical mesh lattice, where the arrival time of optical pulses propagating in a pair of coupled optical fiber loops is interpreted as a synthetic spatial dimension. Here, we develop a general theory of the fluid of light in such optical mesh lattices. On the one hand, this theory provides a solid framework for an analytical and numerical interpretation of the experimental observations. On the other hand it anticipates new physical effects stemming from the specific spatio-temporally periodic geometry of our set-up. Our work opens the way towards the full exploitation of optical mesh lattices system as a promising platform for studies of hydrodynamics phenomena in fluids of light in novel configurations.
	\end{abstract}
	\maketitle

	\section{Introduction}
	
	In recent years, there has been significant growth in the field of ``fluids of light", which aims to bring ideas from quantum gases into nonlinear optics~\cite{RevModPhys.85.299}. Although historically based on analogies between paraxial light propagation in nonlinear systems and the Gross-Pitaevskii equation (GPE) of weakly-interacting atomic gases~\cite{Pomeau1993,Arecchi1991,Swartzlander:PRL1992,Frisch:PRL1992,Staliunas_book}, experiments in this field expanded dramatically in the mid-2000s thanks to the observation of exciton-polariton condensates in semiconductor microcavities~\cite{kasprzak2006bose}. This led to rapid progress in observing superfluidity effects and topological excitations of exciton-polariton fluids~\cite{Amo:NPhys2009,Amo:2011Science,Nardin:2011NatPhys}, and has fuelled wider interest in hydrodynamic effects in light~\cite{liberal2020near}. This interest has also grown even further over the last decade, thanks to developments in artificial magnetic fields for light~\cite{RevModPhys.91.015006} and cavity-less propagating geometries~\cite{leboeuf2010,vocke2015experimental, michel2018superfluid,fontaine2018observation}.
	
	Building on previous experiments with a focus on quantum-optical effects~\cite{schreiber2010photons,regensburger2012parity}, we have recently shown that fluids of light can also be studied in a so-called optical mesh lattice~\cite{experiment}. The central idea of this set-up is to exploit the arrival time of classical optical pulses propagating in coupled optical fiber loops to encode one (or more~\cite{schreiber2,muniz}) discrete synthetic spatial dimensions~\cite{schreiber2010photons,regensburger2012parity,miri2012optical}. In contrast to more traditional platforms for fluids of light~\cite{RevModPhys.85.299,vocke2015experimental,fontaine2018observation,michel2018superfluid}, the flexibility of our set-up allows for the observation of the full light-field dynamics for long propagation times; the engineering of arbitrary dynamical potentials, such as both stationary and moving defects; and the control of the effective nonlinearity in a wide range with standard optoelectronic components. Previously, these advantages have been exploited in studies of, e.g., $PT$-symmetric physics~\cite{miri2012optical, regensburger2012parity, wimmer2013optical,wimmer2015observation, wimmer2015observation:scirep,  muniz} and topological effects~\cite{Chen2018, Chalabi2019,bisianov,weidemann2020topological}. 
	In our recent experiment~\cite{experiment}, we observed several key qualitative signatures of superfluid-like behaviour, including a non-zero speed of sound due to nonlinear effects and the breakdown of apparent superfluidity above a critical-velocity threshold.

	In the present paper, we develop a general and complete theory for fluids of light in optical mesh lattices and we present its application to our experimental set-up. The full development of this theory requires a detailed study of a number of features that are specific to the novel platform, namely the effect of the peculiar connectivity of the optical mesh lattice and the Floquet nature of the evolution given by the repeated propagation of the light pulses around the fiber loops. In addition to providing a solid interpretation of the experimental observations in~\cite{experiment}, we describe new superfluid hydrodynamics effects which will directly follow from the new features of our platform. These include a more complex dependence of the speed of sound on the density, the onset of dynamical instabilities of the superfluid at high densities and a weakened superfluidity by Umklapp processes due to the spatial periodicity.

	The structure of the article is built with two types of readers in mind. Firstly, researchers interested in reproducing the experiment and/or understanding all details of its theoretical interpretation will find that the full article is a reference summarizing all important aspects of optical mesh lattices for hydrodynamic experiments. Secondly, readers that are interested in the general physics of fluids of light in optical mesh lattices can focus on the general theory and discussion of analogies and differences with other platforms for fluids of light, but may skip those subsections that are mostly devoted to the details of realistic experimental set-ups.
	Specifically, in Sec.~\ref{sec:intro}, we present an in-depth analysis of the evolution equations for the nonlinear optical mesh lattice. Building atop the known theory, we present a complete theory of photonic bands in the linear regime and of the Bogoliubov collective excitations in a weakly interacting fluid of light. Analogies with relativistic Dirac-type superfluids are also drawn. In Sec.~\ref{sec:measure}, we numerically demonstrate how to accurately measure the corresponding speed of sound from the effects of a stationary defect in a realistic optical mesh lattice experiment. While Subsec.~\ref{sec:sosnumerical} is of general interest, the following Subsecs.~\ref{sec:real}-\ref{sec:expan} address all those details that are of crucial interest for experiments. This provides a comprehensive theoretical framework to the results previously presented in the main text and the supplemental material of Ref.~\cite{experiment}.
	In Sec.~\ref{sec:landau}, we numerically investigate the effective critical velocity for superfluidity and, in particular, the impact of Umklapp processes due to the periodicity of the optical mesh lattice. To this purpose, we explore a restricted Landau criterion and we identify the apparent threshold for the emission of collective excitations by a moving defect. Once again, the first subsections Subsec.~\ref{sec:analyticslandau}-\ref{sec:numericslandau} are of general interest, with specific experimental details being discussed in Subsec.~\ref{sec:explandau}. Finally, we draw conclusions in Sec.~\ref{sec:conclusions}.

	\section{Nonlinear optical mesh lattices} \label{sec:intro}
	
	In this section, we shall begin by introducing light evolution in nonlinear optical mesh lattices. We then review the derivation of the Bogoliubov dispersion for weak perturbations propagating on top of a stationary and uniform optical field. As we show, this predicts a speed of sound for low-wavelength and low-energy excitations, which tends to zero in the limit of vanishing nonlinearity. As we discuss, this system shares many features with a Dirac-type relativistic superfluid, such as the existence of an instability for high defocusing nonlinearity and a maximum speed of sound. The nonlinear optical mesh lattice therefore opens the way towards the experimental investigation of nonlinear and relativistic effects. 
	
	\subsection{Evolution in a nonlinear optical mesh lattice}  \label{sec:evol}
	
	\begin{figure*}[!]
		\includegraphics[width=0.9\linewidth]{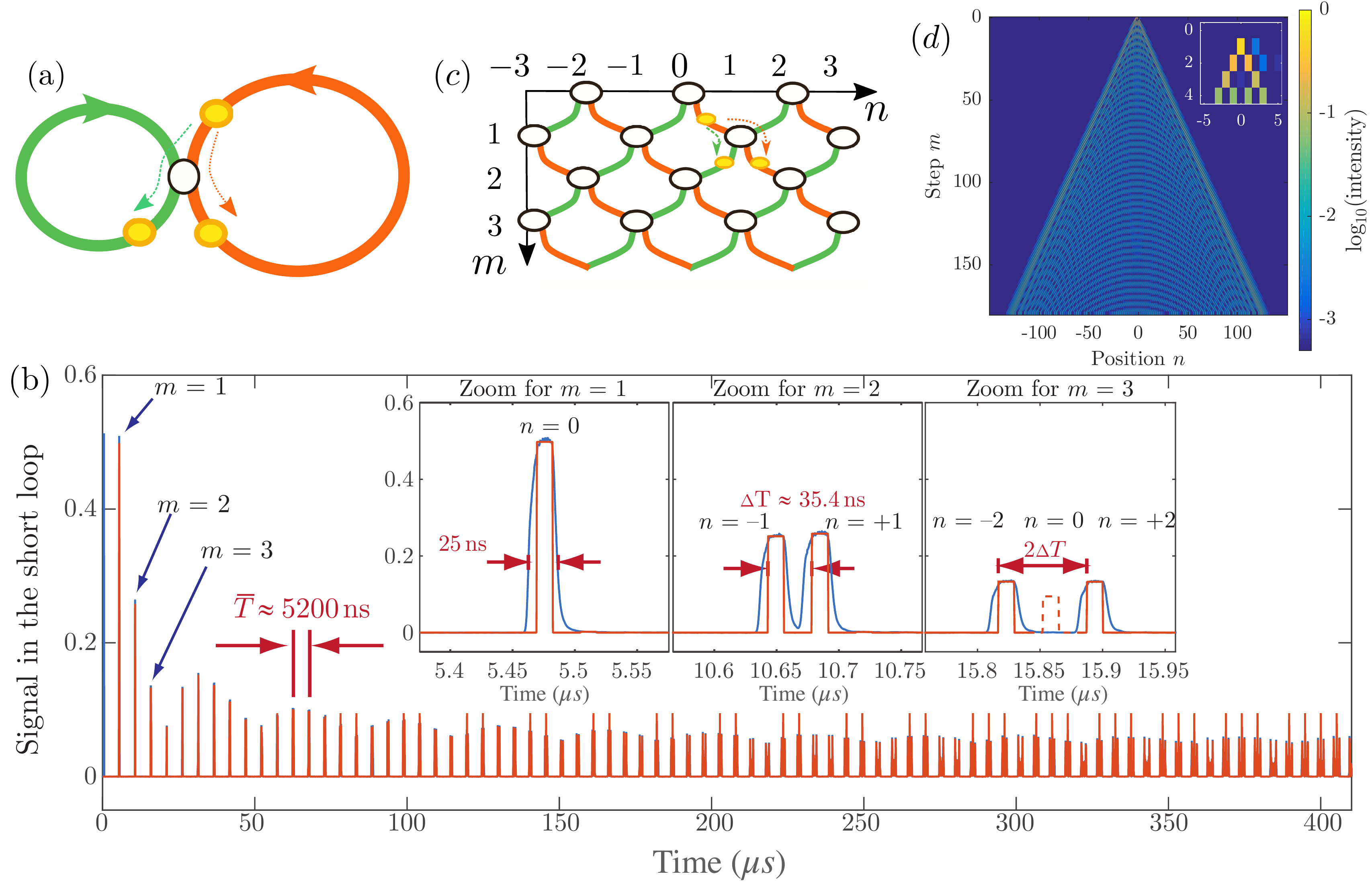}
		\caption{(a) Optical pulses propagate around a long and a short fiber loop, coupled by a 50/50 beamsplitter.  (b) To illustrate the basic operation of this set-up, we plot here the time-dependent light intensity signal ({\it upper blue line}) measured in the short loop  in a simple example of ``Light Walk" experiment, where a single optical pulse is  initially injected into one loop of a linear system~\cite{regensburger2011photon, martinthesis}. Over time,  a train of pulses is observed and the average pulse height of each pulse is extracted ({\it lower orange line}). Amplifiers are present in both loops in order to compensate for losses during the experiment. As can be seen, a distinct group of pulses appears after each time period $\bar{T}$, corresponding to the average round-trip time.  (In this example,  $\bar{T}\approx 5200$ns as the average loop length was roughly 1km.) When we zoom in, we see that pulses within a group are further separated by multiples of $\Delta T$ (here, $\Delta {T}\approx 35.4$ns) corresponding to the time delay for a pulse to travel around the long rather than the short loop. Thanks to a clear separation of the timescales $\bar{T}$ and $\Delta T$, we can label each pulse by ($m,n$), where these are integers counting, respectively, the total number of round trips and the excess number of round trips in the long compared to the short loop. Note that, as can be seen in the inset, for the present ``Light Walk" experiment  there is destructive interference between the different optical paths to $(m,n)=(3,0)$, so that the central peak is absent. (c) A schematic showing the reinterpretation of the two integers, $n$ and $m$, as, respectively, the discrete position in a 1D lattice, $n$, and the discrete time step, $m$. Under this mapping, completing a round-trip in the short or long loop in (a) corresponds respectively to travelling from northeast or northwest to the southwest or southeast in the diamond lattice sketched in (c). (d) The average pulse heights extracted from the example in panel (b) can be replotted in terms of $m$ and $n$ to reveal the time-evolution of the light field in the optical mesh lattice. For this example, the light spreads out with the characteristic distribution of a ``Light Walk", with interference effects clearly being visible. Note that due to the intrinsic diamond connectivity, only either even or odd lattice sites are physically accessible at each time step. Panels (b) and (d) are reproduced from Ref.~\cite{experiment}. }
		\label{overview}	
	\end{figure*}
	
	The discrete time-evolution of light moving in a one-dimensional optical mesh lattice is realised via a time-multiplexing scheme, based on two coupled optical fiber loops~\cite{schreiber2010photons,regensburger2012parity,miri2012optical}, as shown in Fig.~\ref{overview}(a). In the basic set-up, a light pulse is injected into one loop, and it then propagates around the loop until it is split by the ($50/50$) beamsplitter into two pulses, one circulating in each loop. When these pulses again reach the beamsplitter, they are each split into two more pulses and so on. A detector in one of the loops then observes a train of pulses over time, as shown in the example in Fig.~\ref{overview}(b).

	The time-multiplexing scheme relies on choosing the lengths of the loops, $L_1$ and $L_2$, such that there is a small but non-zero length difference, $\Delta L = L_1 - L_2$, as well as a long average length $\tilde{L}= (L_1 + L_2)/2$, with $\tilde{L} \gg \Delta L$. This guarantees that there is  a clear separation of time-scales between the average round-trip time, $\tilde{T}= (T_1+T_2)/2$ and the relative time-delay, $\Delta T = T_1-T_2$, where $T_1$ ($T_2$) are the times taken for a light pulse to travel around the long (short) loop. Provided that the pulse width is narrower than the minimum pulse separation $\Delta T$, the arrival time of the pulses from each loop (before they are combined) at the beamsplitter, can be expressed as $T= m \tilde{T} + n \Delta T/2$, where $m$ and $n$ are two integers. Physically, the integer $m$ counts the total number of round trips, while $n$ counts how many more round trips each pulse made in the long instead of the short loop. This identification is unambiguous provided that all the pulses from each round-trip fit in the time-window $T_2$ set by propagation around the shorter loop; this typically allows for evolutions over several hundred round-trips depending on the length of the fiber loops chosen. At longer times, the identification of individual pulses may not be possible as a pulse with a large positive value of $n$ after $m$ round-trips may overlap with a pulse with a large negative value of $n$ after $m+1$ round-trips. However, experimentally the evolution of the light-field can be confined to small $|n|$ values through the control e.g. of additional amplitude modulators thus realizing absorbing boundary conditions, in which case it is possible to propagate for an even larger number of round-trips~\cite{experiment, martinthesis, alberucci2022wave, muniz2019kapitza}.
	
	To re-interpret the evolution in terms of lattice dynamics, we note that, as the light propagates, the  integer $m$ always increases for each successive round trip, while the integer $n$ can either increase or decrease by one depending on whether the short or long loop is traversed. Guided by this, the integer $m$ can be interpreted as a discrete time step, while $n$ is a discrete position index along a ``synthetic spatial dimension". This leads to the effective $(1+1)$D optical mesh lattice shown schematically in Fig.~\ref{overview}(c). Using this mapping, we can then replot the measured pulse intensities extracted from a time trace [Fig.~\ref{overview}(b)], to reveal the light dynamics in the synthetic optical mesh lattice [see Fig.~\ref{overview}(d)]. 
	
	\begin{figure*}[!]
		\includegraphics[width=0.9\linewidth]{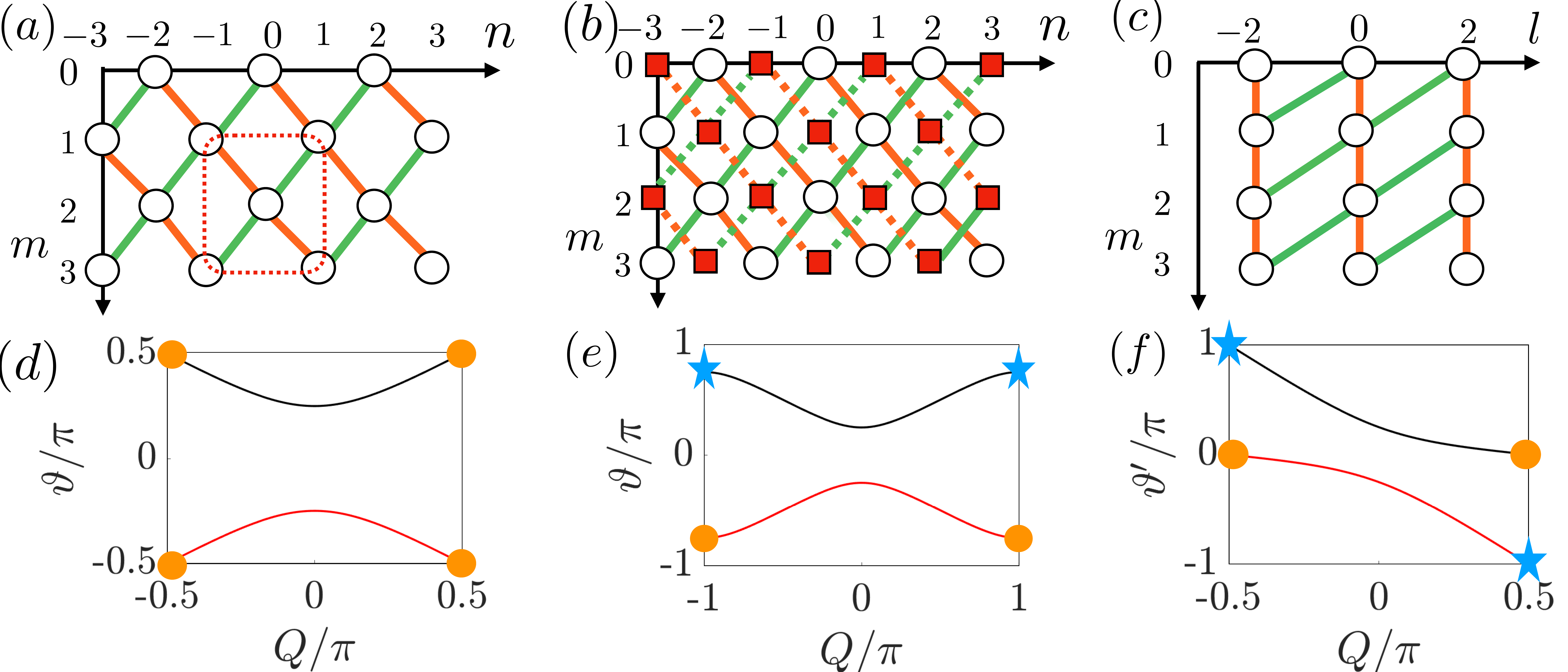}
		\caption{(a) The optical mesh lattice (re-drawn from Fig.~\ref{overview}), with a red dashed box indicating double-steps in time and space, as used in Calculation Method 1. (b) The extended mesh lattice used in Calculation Method 2, with auxiliary sites (red squares) as well as physical sites (white circles). In the extended lattice, auxiliary sites and physical sites are always decoupled due to the diamond connectivity (dotted and solid lines). (c) The optical mesh lattice as viewed in a moving frame in Calculation Method 3, with $l=n-m$. (d)-(f) Plots of the Floquet quasi-energy dispersion in a uniform $\phi_n^m=\varphi_n^m=0$ and linear $\Gamma=0$ optical mesh lattice, as calculated using pictures (a)-(c) respectively. Note that (f) shows the quasi-energy in the moving frame $\theta'$ (for the laboratory frame, see Fig.~\ref{moving}). While these methods all give physically consistent results, each has its own subtleties (see main text). In brief, the varying scales of the axes indicate the different sizes of the first Floquet-Bloch Brillouin zone for each method. At the Floquet-Bloch Brillouin zone boundary, corresponding states are indicated by yellow circles and blue stars, showing significant differences between each method. Firstly, for Method 1 in (d), the two bands cross at the zone boundaries, indicating a missed symmetry. Secondly, for Method 2 in (e), the bands are non-degenerate and do not touch, but the total number of states has doubled as auxiliary degrees of freedom have been included beyond the physical ones. Thirdly, for Method 3 in (f), the upper-band (lower-band) state at $Q=-\pi/2$ is continuously connected to the lower-band (upper-band) state at $Q=\pi/2$, indicating an unusual boundary condition (see Fig.~\ref{moving}). 
		}\label{fig:methods}	
	\end{figure*}

	In all experiments pulses were long enough and the loop lengths similar enough that dispersion effects do not play a significant role for the experimentally realised propagation length: the pulses emerging after each round-trip have an almost identical shape independently of their previous path along short and long loops, which guarantees their perfect overlap at the beam-splitter. Under these assumptions, each pulse can be characterized by a single complex amplitude only, as further discussed e.g. in Refs.~\cite{martinthesis, wimmer2013optical}. The pulse distribution at step $m+1$ is then described by the evolution equations~\cite{regensburger2011photon,Navarrete,wimmer2013optical}:
	\begin{eqnarray}
		u_n^{m+1} &=& \frac{1}{\sqrt{2}} \left( u_{n+1}^m e^{i \Gamma |u_{n+1}^m|^2} + i v_{n+1}^m e^{i \Gamma |v_{n+1}^m|^2} \right) e^{i \varphi_n^m} , \label{eq:nonlinear_1} \\
		v_n^{m+1} &=& \frac{1}{\sqrt{2}} \left( v_{n-1}^m e^{i \Gamma |v_{n-1}^m|^2} + i u_{n-1}^m e^{i \Gamma |u_{n-1}^m|^2} \right) e^{i \phi_n^m}, \label{eq:nonlinear_2}
	\end{eqnarray}
	where the two components $u_n^m$ and $v_n^m$ of the wavefunction denote the pulse amplitude entering the beamsplitter from the short and long loop, respectively. The relative phase shift of $\pi/2$ for light that couples from one loop to the other loop is a consequence of energy conservation at the symmetric $50/50$ beamsplitter. Note that an unusual feature of the optical mesh lattice is its diamond connectivity over time, as physically, light either travels through the short loop ($n \rightarrow n-1$) or the long loop ($n \rightarrow n+1$). 
	
	To reach the nonlinear optical regime, dispersion compensating optical fibers can be used for the fiber loops, as these have a significantly higher nonlinearity and suppress the growth of modulational instabilities inside the pulses when the light propagates over long distances with high intensity~\cite{agrawal2013nonlinear}.
	In our formalism, the coefficient $\Gamma$ accounts for the effective nonlinear action accumulated within  a single round trip, and is taken to be the same for both loops. Note that a positive $\Gamma$ (as describes usual dispersion-compensating fibers) actually corresponds to having a negative interaction energy in the language of quantum fluids. 
	
	The phase-shifts $\varphi_n^m$ and $\phi_n^m$ are controlled by phase-modulators inserted in each loop. As these modulators can respond faster than the minimum pulse separation, $\Delta T$, the imposed phase shifts can be designed with an arbitrary dependence on  both pulse indices, $m$ and $n$. For particles moving in a lattice, this is analogous to adding an effective potential, which varies both with position and time step, and which can moreover be different for the two components $u_n^m,v_n^m$ of the wave-function. This provides a versatile way to control the properties of optical mesh lattices; for example, previous experiments have utilised specially-designed phase together with amplitude modulations as a tool to create and investigate $\mathcal{PT}$-symmetric optical mesh lattices~\cite{regensburger2012parity, regensburger:2013, wimmer2015observation,wimmer2015observation:scirep}, to investigate optical diametric drive acceleration~\cite{wimmer2013optical}, to engineer and map out the Berry curvature of the optical band-structure~\cite{wimmer2017experimental} and to confine (i.e. trap) the optical field around a desired position, $n$~\cite{experiment}. In this work, we use the phase-shifts as a way to imprint defects on the light field in Sec.~\ref{sec:measure} and Sec.~\ref{sec:landau}.

	\subsection{Photonic band structure}
	\label{sec:band}
	
	In order to physically understand the behaviour of the mesh lattice, the first step of our study is to look for the band structure for vanishing nonlinearity $\Gamma=0$ and vanishing phases $\varphi_n^m=\phi_n^m=0$. However, there are important subtleties in this derivation stemming from the unusual diamond connectivity of the lattice with respect to the discrete time step (Fig.~\ref{overview}(b)). To illustrate this, we shall now discuss, in turn, three distinct theoretical approaches, based on introducing, respectively: (1) double steps in time and space, (2) an extended lattice with auxiliary lattice sites and (3) a moving frame. Each of these approaches gives physically equivalent results, but with subtleties that need to be taken into account. 
	
	\subsubsection*{Calculation method 1: Double steps}
	
	Given the diamond connectivity shown in Fig.\ref{overview}(c), the most simple approach is to anticipate that the optical mesh lattice is periodic under double steps in the discrete time step $m\rightarrow m+2$ and the position $n\rightarrow n+2$. The linear evolution equations for a double time step from \eqref{eq:nonlinear_1}-\eqref{eq:nonlinear_2} are:
	\begin{eqnarray}
		u_n^{m+2} &=& \frac{1}{{2}} \left( u_{n+2}^m  + i v_{n+2}^m + i v_{n}^m -u_n^m \right)  ,  \\
		v_n^{m+2} &=& \frac{1}{{2}} \left( v_{n-2}^m  + i u_{n-2}^m + i u_{n}^m -v_n^m \right) , 
	\end{eqnarray}
	where we have set $\Gamma=0$ and $\varphi_n^m=\phi_n^m=0$. 
	One can then look for plane-wave solutions of the form~\cite{regensburger2012parity, miri2012optical}:
	\begin{eqnarray}
		\left( \begin{array} {c} u^m_n \\ v^m_n \end{array} \right)= \left( \begin{array} {c}\bar{u} (Q)\\\bar{v} (Q) \end{array} \right) e^{i Q n} \, e^{-i \vartheta m}, \label{eq:ansatz}
	\end{eqnarray}
	with wavevector $Q$ and Floquet quasi-energy $\vartheta$. Insertion of this Floquet-Bloch ansatz into the double-step equations, leads to the characteristic equation for the band structure 
	\begin{equation}
		e^{-2 i\vartheta} \left( \begin{array}{c} \bar{u} \\ \bar{v} \end{array}\right) = \frac{1}{{2}} \left(  \begin{array}{cc} e^{2iQ} -1 & i (e^{2iQ} +1) \\ i  (e^{-2iQ} +1) & e^{-2iQ} -1 \end{array} \right)  \left(  \begin{array}{c} \bar{u} \\ \bar{v}\end{array} \right) ,
	\end{equation}
	which can be solved to give the dispersion relation~\cite{miri2012optical}:
	\begin{equation}
		\cos \vartheta_j  = \frac{1}{\sqrt{2}} \cos Q ,    \label{eq:linear_dispersion}
	\end{equation}
	where $j=\pm 1$ is the band index and the wavevector $Q$ and Floquet quasi-energy $\vartheta$ are defined within the Floquet-Brillouin zone $-\pi/2 \leq Q,\vartheta < \pi/2$, reflecting that we are considering double steps in both position and time. This dispersion is plotted in Fig.~\ref{fig:methods}(d); however, as can be seen, there is a band-crossing (indicated by yellow circles) at the Floquet-Brillouin zone boundary (i.e. at $|Q|=|\vartheta|=\pi/2$), which hints at some missed symmetry in this method.  This remaining symmetry can be broken and the band-degeneracy is lifted if we impose explicit double-step modulations in the evolution, for example through the phases, $\varphi_n^m, \phi_n^m$. In such cases, an even cleaner approach can be used to redefine $m$ as the number of double round-trips, such that the light evolution is described by a set of four coupled equations (instead of two). However, as we do not consider explicit double-step modulations in this paper, we will focus on other calculation approaches, which do not give these spurious band-crossings. 
	
	\subsubsection*{Calculation method 2: Auxiliary lattice}
	\label{sec:auxiliary}
	
	The second method we will discuss is to artificially extend the lattice by adding in auxiliary sites (red squares) at odd (even) positions at even (odd) time steps, as shown in Fig.\ref{fig:methods}(b). When taken together with the physical lattice sites (white circles), this means that we have a full square lattice (in terms of discrete time-position space), except with only diagonal connectivity (dotted and solid lines). Thanks to this 
	connectivity, a physical field which is initialized at $m=0$ with a non-vanishing amplitude only in the physical even sites will continue to only be non-zero on physical even (odd) sites at all even (resp. odd) time steps $m$. 
	
	Since the system of physical plus auxiliary sites is a square lattice, we can apply the ansatz~\eqref{eq:ansatz} to the single-step evolution equations~\eqref{eq:nonlinear_1}-\eqref{eq:nonlinear_2}, taking $\Gamma=0$ and $\varphi_n^m=\phi_n^m=0$. The corresponding characteristic equation for the band structure is
	\begin{equation}
		e^{-i\vartheta} \left( \begin{array}{c} \bar{u}_j(Q) \\ \bar{v}_j(Q) \end{array}\right) = \frac{1}{\sqrt{2}} \left(  \begin{array}{cc} e^{iQ} & i e^{iQ} \\ i e^{-iQ} & e^{-iQ} \end{array} \right)  \left(  \begin{array}{c} \bar{u}_j(Q) \\ \bar{v}_j(Q) \end{array} \right) \label{eq:auxiliary}
	\end{equation}
	which gives the same form for the dispersion as Eq.~\ref{eq:linear_dispersion}, except now defined within $-\pi\leq\vartheta,Q<\pi$ as plotted in Fig.~\ref{fig:methods}(e). Note that the periodicity of the dispersion in the Bloch momentum and the Floquet quasi-energy is now consistent with the unit step periodicity of the lattice in both spatial and temporal directions and there are no spurious band crossings. 
	
	However, as we still find the same form for the dispersion despite having doubled the domain of $Q$ (from $-\pi/2\leq Q<\pi/2$ in Method 1 to $-\pi\leq Q<\pi$ in Method 2), we have actually found twice as many states as before. This can be understood by remembering that we artificially doubled our degrees of freedom by adding in the nonphysical auxiliary lattice sites. Therefore, while the dispersion relation for this extended lattice appears to be simple, the condition that the field is non-zero only on the physical sites (while vanishing on the auxiliary sites) must be reflected in the momentum-space picture in a subtle way. In fact, it is not encoded in the band dispersion (or in the selection of a particular branch), but rather it can be expressed in terms of a restriction on the amplitude of the field in the different Bloch modes of the extended lattice. To see this, we note from Eq.~\ref{eq:auxiliary} that the eigenmodes satisfy the symmetry condition
	\begin{equation}
		\left( \begin{array}{c} \bar{u}_j(Q) \\ \bar{v}_j(Q) \end{array}\right) = \left( \begin{array}{c} \bar{u}_{-j}(Q+\pi) \\ \bar{v}_{-j}(Q+\pi) \end{array}\right). \label{eq:symm}
	\end{equation}
	It is then easy to verify that the momentum-space field amplitudes of a generic physical initial state must satisfy 
	\begin{equation}
		\alpha_j(Q)=\alpha_{-j}(Q+\pi), \label{eq:physical_state}
	\end{equation}
	in order that the field vanishes on the auxiliary sites, which were only introduced for mathematical convenience. The relation $\theta_{-j}(Q+\pi)=\theta_j(Q)+\pi$ between the Floquet quasi-energies of the two bands guarantees that the condition \eqref{eq:physical_state} remains fulfilled at arbitrary even time steps $m$ during the time-evolution. At odd time steps, instead, this same condition guarantees that the field amplitude is non-vanishing only at odd sites (i.e. the physical lattice sites).
	
	As Eq.\eqref{eq:physical_state} implies, any physical state must contain pairs of Bloch eigenstate of the extended lattice, and any generic physical state will contain several Floquet quasi-energies. This pairing of states removes the unphysical auxiliary degrees of freedom, making the state-counting for physical states consistent again with that in Method 1. As an explicit example of this state-pairing, let us consider the concrete example of a field that is uniform throughout the whole physical 1D lattice (i.e. that at, an even time step, has an equal amplitude at all even positions and zero amplitude at all odd positions). In terms of the Bloch eigenstates of the extended lattice, this corresponds to the superposition of the two states $Q=0$, $j=1$ and $Q=\pi$, $j=-1$ for the upper band or of the two states $Q=0$, $j=-1$ and $Q=\pi$, $j=1$ for the lower band.

	\begin{figure}[!]
		\includegraphics[width=0.99\linewidth]{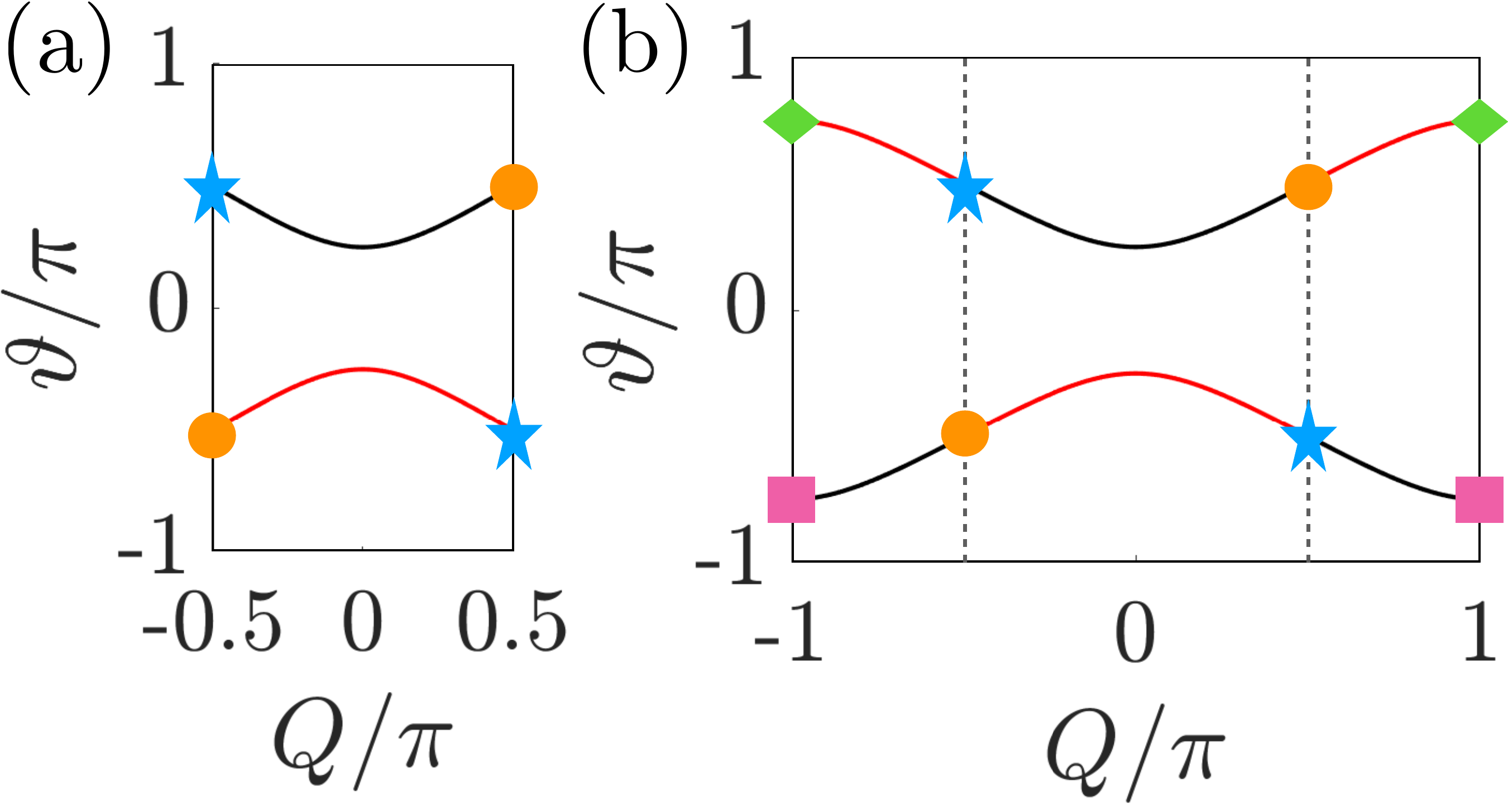}
		\caption{(a) The Floquet quasi-energy dispersion obtained after transforming the dispersion in Fig.~\ref{fig:methods}(f) for the moving lattice (Method 3) back into the laboratory frame via the Galilean transform  $\theta = \theta' +Q$. The blue stars and yellow circles indicate the correspondence between states identified in Fig.~\ref{fig:methods}(f). (b) Applying these boundary conditions, we can construct the dispersion in an extended-zone between $-\pi \leq Q < \pi$ with the first  Brillouin zone marked by vertical grey dotted lines. This is fully equivalent to the dispersion obtained via Method 2, shown in Fig.~\ref{fig:methods}(c), and the colours of the curves can be seen as representing the symmetry condition noted in~\eqref{eq:symm}. With respect to this larger momentum range, the dispersion is periodic and has normal boundary conditions at the zone boundaries (as indicated by green diamonds and pink squares). 
		}\label{moving}	
	\end{figure}
	
	Introducing auxiliary lattice sites is therefore a useful and simple method for carrying out calculations, but care is required when interpreting the results, as we shall return to again when discussing Bogoliubov calculations in the next section. We also note that as the specific connectivity of the extended lattice  keeps the physical and unphysical auxiliary sectors totally disconnected, there is no practical problem in restricting our attention to one of the two components only and keeping in mind only at the end that the physical field has to contain an implicit multiplication by $\textrm{mod}(n+m,2)$. 
	
	\subsubsection*{Calculation method 3: A moving frame} \label{sec:moving}
	
	The final approach we shall discuss is to carry out the calculation in a quasi-moving frame [see Fig.~\ref{fig:methods}(c)]. Here we fix the position zero to multiples of the round trip time of the long loop $T_1$ and not to the average round-trip time $\tilde{T}$ as was done before. This avoids the difficulties of the two previous methods but introduces a new subtlety. To transform into this quasi-moving frame, we redraw the lattice and change the coordinate system to $(m, l)$ where $l\!= n\!-\!m\!$. In this frame, the nonlinear equations (Eqs.~\ref{eq:nonlinear_1} and~\ref{eq:nonlinear_2}) with vanishing phases $\varphi_n^m=\phi_n^m=0$ become:
	\begin{eqnarray}
		u_{l}^{m+1} &=& \frac{1}{\sqrt{2}} ( u_{l+2}^m e^{i \Gamma |u_{l+2}^m|^2} + i v_{l+2}^m e^{i \Gamma |v_{l+2}^m|^2} ) \label{eq:nonlinearmoving1} \\
		v_{l}^{m+1} &=& \frac{1}{\sqrt{2}} ( v_{l}^m e^{i \Gamma |v_{l}^m|^2} + i u_{l}^m e^{i \Gamma |u_{l}^m|^2} )  \label{eq:nonlinearmoving2}
	\end{eqnarray}
	and so we only need to consider even values of $l$ in our lattice. Considering the linear case ($\Gamma=0$), we can solve these equations by using the Floquet-Bloch ansatz~\eqref{eq:ansatz}, making the replacement $\vartheta \rightarrow \vartheta'$, as we are now solving for the quasi-energy $\vartheta'$ in the moving frame. The characteristic equation for the band structure is then
	\begin{equation}
		e^{-i\vartheta'} \left( \begin{array}{c} \bar{u}_l(Q) \\ \bar{v}_l(Q) \end{array}\right) = \frac{1}{\sqrt{2}} \left(  \begin{array}{cc} e^{2iQ} & i e^{2iQ} \\ i  & 1 \end{array} \right)  \left(  \begin{array}{c} \bar{u}_l(Q) \\ \bar{v}_l(Q) \end{array} \right) \nonumber
	\end{equation}
	which leads to the dispersion 
	\begin{equation}
		\cos \left(\vartheta_j'+Q \right)  = \frac{1}{\sqrt{2}} \cos Q ,\label{eq:moving}
	\end{equation}
	as plotted in Fig.~\ref{fig:methods}(f). Note now $-\pi/2\leq Q<\pi/2$ and $-\pi\leq\vartheta'<\pi$, as our equations~\eqref{eq:nonlinearmoving1}-\eqref{eq:nonlinearmoving2} describe two steps along $l$ and one step along $m$. As can be seen, the dispersion has two slanted bands as it is in a moving frame. 
	
	To transform back into the laboratory frame, we can make use of a Galilean transform $\vartheta = \vartheta' +Q$ where the moving frame is moving at a speed of one site per time step, as can be seen from Fig.~\ref{fig:methods}(c). Then from Eq.~\ref{eq:moving} we recover the dispersion~\eqref{eq:linear_dispersion} found in Methods 1 and 2, except now with the Bloch momentum defined over $-\pi/2\leq Q <\pi/2$ and the quasi-energy over $-\pi\leq \vartheta <\pi$ as shown in Fig.~\ref{moving}(a). This confirms that Method 3 is physically consistent with other methods, but with the advantage of avoiding the issues with spurious band-crossings and over-counting of states. 
	
	However, Method 3 does have its own important subtlety, as there are now unusual boundary conditions at the Brillouin zone boundary stemming from the Galilean transform. As can be seen comparing Fig.~\ref{fig:methods}(f) and Fig.~\ref{moving}(a), we must identify the upper (lower) band eigenstate at $Q=-\pi/2$ with the lower (upper)  band eigenstate at $Q=\pi/2$. In fact, this is the same symmetry condition as~\eqref{eq:symm}, and corresponds to a shift of $\pi$ in the quasi-energy when we match-up opposite edges of this Brillouin zone. This becomes crucial when we use the laboratory-frame dispersion [Fig.~\ref{moving}(a)] to construct the extended-zone dispersion in Fig.~\ref{moving}(b), corresponding to the first two Brillouin zones. Here, we see that by using the unusual boundary conditions we recover the same dispersion as in Method 2 [Fig.~\ref{fig:methods}(e)]. An advantage of using Method 3 as compared to Method 2 is that the symmetry condition on the eigenstates~\eqref{eq:symm} is  immediately clear from the colouring of the bands. Note also that for this larger momentum range, $-\pi \leq Q < \pi$, we recover normal boundary conditions, as shown by green diamonds and pink squares in Fig.~\ref{moving}(b). 
	
	\begin{figure*}[!]
		\includegraphics[width=0.99\linewidth]{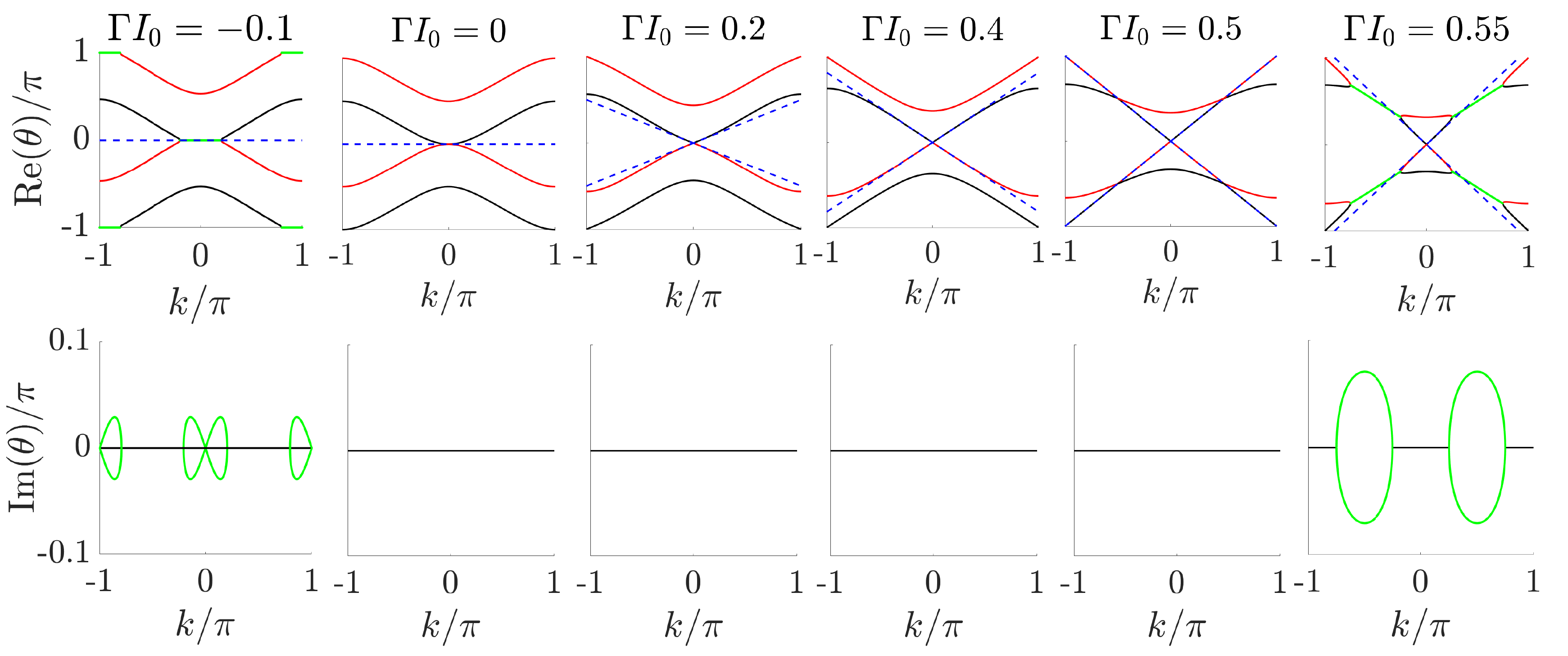}
		\caption{The real ({\it top panels}) and imaginary ({\it bottom panels}) parts of the Bogoliubov dispersion for an unperturbed light field in the lower band of the nonlinear optical mesh lattice (Eq.~\ref{eq:bog}), for different values of the nonlinearity parameter:  $\Gamma I_0 =-0.1, 0, 0.2, 0.4, 0.5, 0.55$ ({\it left to right}). The phonon-like linear dispersions (Eq.~\ref{eq:so}) are marked by the blue dashed lines. The black/red/green ({\it in greyscale resp. black/grey/light grey}) color of each curve indicates the negative/positive/zero value of the Bogoliubov norm of the band, respectively. For $\Gamma I_0<0$ and $\Gamma I_0>0.5$, the bands develop a non-zero imaginary part, indicating dynamical instability.}\label{bogdispersion}	
	\end{figure*}

	\subsection{Bogoliubov dispersion} \label{sec:bog}

	In order to derive the Bogoliubov dispersion of collective excitations, one needs to consider the dynamics of  small perturbations, $\delta u_n^m$ and $\delta v_n^m$, on top of a stationary and initially unperturbed light field, i.e. such that $u_n^m \rightarrow  u_n^m + \delta u_n^m$ and $v_n^m \rightarrow  v_n^m + \delta v_n^m$~\cite{pitaevskii2016bose,CastinLectures}. We shall now show how the Bogoliubov dispersion can be obtained by extending either the above Method 2 (of auxiliary lattice sites) or the above Method 3 (of a moving frame).

	\begin{figure}[!]
		\includegraphics[width=0.83\linewidth]{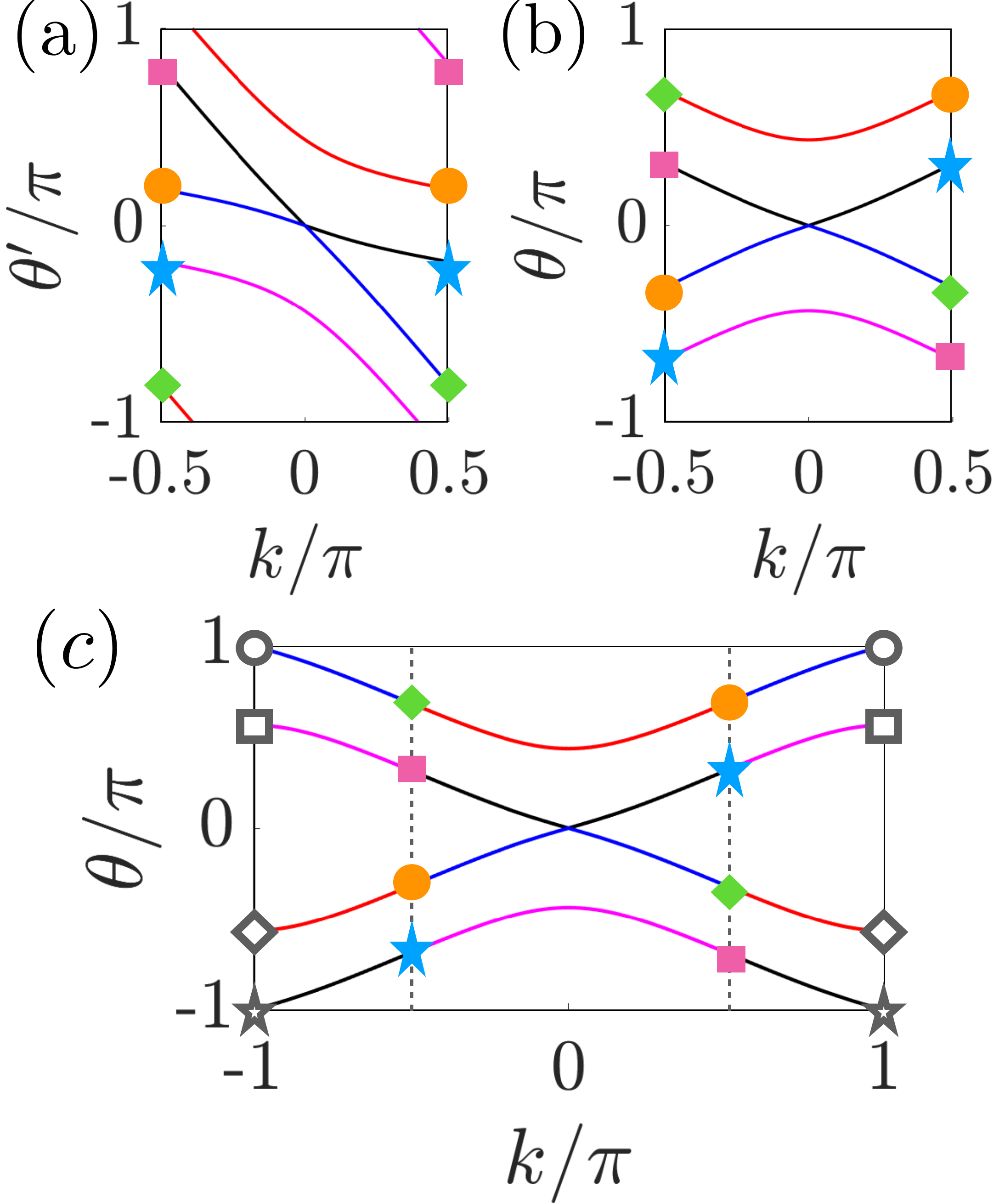}
		\caption{(a) The Bogoliubov quasi-energy dispersion~\eqref{eq:bogmoving} for $\Gamma I_0=0.2$ in the moving frame. The continuous connection of the bands across the first Brillouin zone boundaries with the  unusual boundary conditions of this Method (see also Fig.~\ref{moving}) are indicated by the symbols; bands with $(j, \eta) = (-1, -1)$, $(-1, 1)$, $(1, -1)$, $(1, 1)$ are colored in pink, blue, black and red respectively (as ordered from the bottom to the top at $k=0$). (b) The dispersion of (a) Galilean-transformed back into the laboratory frame via  $\theta =\theta' +k$. (c) The extended zone scheme between $-\pi \leq k < \pi$ constructed from (b) (with the first  Brillouin zone marked by vertical grey dotted lines). This is fully equivalent to the Bogoliubov dispersion obtained via Method 2, shown in Fig.~\ref{bogdispersion}, with the colours of the curves now also showing the the symmetry condition on the states. With respect to this larger momentum range, the dispersion is periodic and has normal boundary conditions (as indicated by open grey symbols).
		}\label{movingbog}	
	\end{figure}
	
	\subsubsection*{Bogoliubov theory from Method 2}
	
	Continuing from Sec.~\ref{sec:auxiliary}, we assume the unperturbed light field is at $Q=0$, and is  described by: 
	\begin{eqnarray}
		\left( \begin{array}{c} u_n^m \\ v_n^m \end{array} \right)
		=\left( \begin{array}{c} \sqrt{I_0} \\ \mp \sqrt{I_0} \end{array} \right) e^{\mp i  \pi /4 m} e^{i \Gamma I_0 m}, \label{eq:light}
	\end{eqnarray}
	where $I_0$ is the light intensity in each loop. The two signs correspond to the eigenstates of the $j=\pm 1$ upper and lower bands, and we have inserted the form of the corresponding eigenstates and quasi-energies $\vartheta = \pm \pi/4$ at Q=0 (see Eq.~\ref{eq:linear_dispersion})~\cite{martinthesis}. At $Q=0$, the group velocity of the linear dispersion relation vanishes, and hence the light field is stationary, while the effective mass, $m^*\!\propto\!(\partial^2 \vartheta / \partial Q^2)^{-1}$, is maximal and of opposite (positive/negative) sign in the two (upper/lower) bands.
	
	Note that this calculation considers a field that is non-vanishing at all sites and implicitly assumes that the physical field is obtained after multiplying by $\textrm{mod}(n+m,2)$. If one wished to restrict the field to the physical sites at all steps, one would have to develop a Bogoliubov theory around a time-dependent state involving two components at different Floquet quasi-energy, which is a much more complicated task. However, since the dynamics of the physical and auxiliary sites are decoupled also at nonlinear level, we anticipate that the Bogoliubov dispersion is not affected by the inclusion of the auxiliary sites.
	
	In the nonlinear regime, the sign of the effective mass has important consequences on the stability of the system~\cite{Morandotti:2001}. In particular, if the nonlinearity, $\Gamma$, and the effective mass, $m^*$, have the same sign, it corresponds to a focusing nonlinearity which destabilises the system, as for a BEC with attractive interactions~\cite{pitaevskii2016bose}. To avoid this instability, a defocusing nonlinearity is required; when the nonlinearity $\Gamma >0$ ($\Gamma <0$), this can be achieved by exciting the lower (upper) band eigenstate, for which the effective mass is negative (positive). From here on, we focus on the $\Gamma >0$ case that is relevant to the experiments~\cite{experiment}, and so only discuss the lower-band eigenstate in~\eqref{eq:light}. 
	
	The small perturbations, $\delta u_n^m$ and $\delta v_n^m$, are described by a temporally and spatially periodic ansatz~\cite{experiment}:
	\begin{eqnarray}
		\left( \begin{array}{c} \delta u_n^m \\  \delta v_n^m \end{array} \right)
		=\left( \begin{array}{c} A_u e^{-i (\theta m - k n)} + B_u^* e^{i (\theta m - k n)} \\ A_v e^{-i (\theta m - k n)} + B_v^* e^{i (\theta m - k n)} \end{array} \right) \label{eq:bogansatz}
	\end{eqnarray}
	where $A_u,A_v,B_u,B_v$ are constant coefficients and where $\theta$ and $k$ are now the ``energy" and ``Bloch momentum" of the Bogoliubov excitations. 
	
	Inserting this ansatz into the nonlinear evolution equations~(\ref{eq:nonlinear_1}-\ref{eq:nonlinear_2}) and linearizing in the small perturbation, one gets the following matrix equation for the collective eigenmodes,
	\begin{equation}
		e^{-i \theta} \vec{\Phi} = \mathcal{L}_k \vec{\Phi}
	\end{equation}
	with $\vec{\Phi}=(A_u,A_v,B_u,B_v)^T$ and the Bogoliubov matrix
	\begin{equation}
		\mathcal{L}_k = \left( \begin{array}{cccc} 
			\frac{1 + i \Gamma I_0}{1 + i} e^{i k} 
			& 
			\frac{i - \Gamma I_0}{1 + i} e^{i k} 
			&
			\frac{i \Gamma I_0}{1 + i} e^{i k}  
			&
			\frac{-  \Gamma I_0}{1 + i} e^{i k} 
			\\
			\frac{i - \Gamma I_0}{1 + i} e^{-i k} 
			& 
			\frac{1 +i \Gamma I_0}{1 + i} e^{-i k} 
			&
			\frac{- \Gamma I_0}{1 + i} e^{-i k}  
			&
			\frac{i \Gamma I_0}{1 + i} e^{-i k}     
			\\
			\frac{- i \Gamma I_0}{1 - i} e^{i k} 
			& 
			\frac{- \Gamma I_0}{1 - i} e^{i k} 
			&
			\frac{1- i\Gamma I_0}{1 - i} e^{i k}  
			&
			\frac{-i - \Gamma I_0}{1 - i} e^{i k}        
			\\
			\frac{-  \Gamma I_0}{1 - i} e^{-i k} 
			& 
			\frac{- i\Gamma I_0}{1 - i} e^{-i k} 
			&
			\frac{-i- \Gamma I_0}{1 - i} e^{-i k}  
			&
			\frac{1 - i\Gamma I_0}{1 - i} e^{-i k}     
		\end{array} \right).\nonumber
	\end{equation}
	Solving this equation leads to the Bogoliubov dispersion relation: 
	\begin{eqnarray}
		\cos \theta &&= \frac{1}{2} \Gamma I_0 \cos k + \frac{1}{2} \cos k \nonumber \\
		&&\pm \frac{1}{2} \sqrt{(\Gamma^2 I_0^2 + 2 \Gamma I_0 + 1) \cos^2 k + 2 \sin^2 k - 4 \Gamma I_0} , \qquad. \label{eq:bog}
	\end{eqnarray}
	which consists of four branches, within the Floquet-Brillouin zone $-\pi \leq k, \theta < \pi$. In Fig.~\ref{bogdispersion}, these branches are plotted in black/red/green depending on the negative/positive/zero value of the Bogoliubov norm $|A_u|^2+|A_v|^2-|B_u|^2-|B_v|^2$. As usual, branches are organized in pairs with opposite quasi-momentum $k$ and Floquet quasi-energy and opposite norm.
	As a consequence of the overall phase-rotation symmetry of the problem, a pair of positive- and negative-norm branches must go through the $k=0$ and $\theta=0$ point according to Goldstone's theorem, and as can be seen here. 
	
	However, as we are implicitly adding auxiliary lattice sites and working in the extended-lattice picture [see Fig.~\ref{fig:methods}(b)], we have again included unphysical degrees of freedom in this calculation. Analogously to Sec.~\ref{sec:auxiliary}, there is a symmetry between states and quasi-energies, e.g. $\vec{\Phi}_ {-j, \eta}  (k+ \pi) = \vec{\Phi}_ {j, \eta}  (k)$ and $\theta_{-j, \eta}  (k+ \pi)= \theta_{j, \eta}(k) + \pi$, where $\eta$ is the sign of the Bogoliubov norm [such that the band labelling goes from bottom to top in Fig.~\ref{bogdispersion} as $(j, \eta) = (-1, -1)$, $(-1, 1)$, $(1, -1)$, $(1, 1)$]. %
	Again, we can circumvent these complications by instead carrying out the calculation in a moving frame. 
	
	\subsubsection*{Bogoliubov theory from Method 3}
	
	Now, following on from Sec.~\ref{sec:moving}, we apply Bogoliubov theory to the moving frame. As above, the unperturbed light field is assumed to be at $Q=0$ on the lower band and described by~\eqref{eq:light}. The small perturbations, $\delta u_l^m$ and $\delta v_l^m$, are also again described by~\eqref{eq:bogansatz} when we rewrite $\theta \rightarrow \theta'$ to represent the ``energy" of the Bogoliubov excitations in the moving frame. Inserting this ansatz into the evolution equations~(\ref{eq:nonlinearmoving1}-\ref{eq:nonlinearmoving2}) and linearizing in the small perturbation, one gets the following Bogoliubov matrix
	\begin{equation}
		\mathcal{L}_k\! =\! \left( \begin{array}{cccc} 
			\frac{1 + i \Gamma I_0}{1 + i} e^{2 i k} 
			& 
			\frac{i - \Gamma I_0}{1 + i} e^{2 i k} 
			&
			\frac{i \Gamma I_0}{1 + i} e^{2 i k}  
			&
			\frac{-  \Gamma I_0}{1 + i} e^{2 i k} 
			\\
			\frac{i - \Gamma I_0}{1 + i} 
			& 
			\frac{1 +i \Gamma I_0}{1 + i}  
			&
			\frac{- \Gamma I_0}{1 + i}   
			&
			\frac{i \Gamma I_0}{1 + i}     
			\\
			\frac{- i \Gamma I_0}{1 - i} e^{2 i k} 
			& 
			\frac{- \Gamma I_0}{1 - i} e^{2 i k} 
			&
			\frac{1- i\Gamma I_0}{1 - i} e^{2 i k}  
			&
			\frac{-i - \Gamma I_0}{1 - i} e^{2 i k}        
			\\
			\frac{-  \Gamma I_0}{1 - i}
			& 
			\frac{- i\Gamma I_0}{1 - i} 
			&
			\frac{-i- \Gamma I_0}{1 - i}   
			&
			\frac{1 - i\Gamma I_0}{1 - i}      
		\end{array} \right).\nonumber
	\end{equation}
	which leads to 
	\begin{eqnarray}
		\cos&& (\theta' +k) = \frac{1}{2} \Gamma I_0 \cos k + \frac{1}{2} \cos k \nonumber \\
		&&\pm \frac{1}{2} \sqrt{(\Gamma^2 I_0^2 + 2 \Gamma I_0 + 1) \cos^2 k + 2 \sin^2 k - 4 \Gamma I_0} ,\quad \label{eq:bogmoving}
	\end{eqnarray}
	which describes four slanted Bogoliubov bands, as shown in Fig.~\ref{movingbog}(a) for $\Gamma I_0=0.2$. 
	Again, this dispersion can be transformed back to the laboratory frame by substituting $\theta=\theta'+k$, recapturing Eq.~\ref{eq:bog} and as shown in Fig.~\ref{movingbog}(b). As before, the major differences with Method 2 are that now $-\pi/2 \leq k < \pi/2$ and $-\pi \leq \theta < \pi$, meaning that Method 3 only captures the desired physical states, but with the price of an unusual boundary condition that connects bands with opposite $j$ (and the same Bogoliubov norm $\eta$) across the Brillouin zone boundary (as shown by symbols in Fig.~\ref{movingbog}). 
	
	Enforcing this boundary condition when plotting an extended-zone scheme in Fig.~\ref{movingbog}(c) then exactly recovers the curves obtained via Method 2 in Fig.~\ref{bogdispersion}. The symmetry that relates eigenstates in different branches can also be clearly visualised in Fig.~\ref{movingbog}(c) through the matching colours of the curves. Within this extended-zone scheme (corresponding to the first two Brillouin zones of the physical lattice), the Bogoliubov dispersion is periodic in the usual sense, as shown by open grey symbols in Fig.~\ref{movingbog}(c). For this reason, hereafter, we shall focus on the Bogoliubov dispersion between $-\pi \leq k < \pi$.   
	
	Through this extended theoretical discussion, we have therefore established the basis for a Bogoliubov description of fluids of light in optical mesh lattices; hereafter we will focus on the physical implications of these results, without further detailed discussion of the underlying subtleties. 
	
	\subsection{Bogoliubov instabilities and speed of sound} \label{sec:bog2}
	
	As can be seen in Fig.~\ref{bogdispersion},
	for finite nonlinearities, the Bogoliubov dispersion~\eqref{eq:bog} can develop non-zero imaginary parts, indicating parameters for which the system is unstable. Firstly, this occurs when $\Gamma <0$, corresponding to having a focusing nonlinearity for the lower-band eigenstate with negative effective mass, as already discussed above. Secondly, and more interestingly, the system is also unstable for a defocusing nonlinearity when $\Gamma > 0.5$. This is a peculiar feature of our two component model and is completely absent in the one-component GPE. It arises due to level attraction between the two opposite-norm Bogoliubov branches in the neighborhood of the two points $k=\pm \pi/2$, leading to exceptional points in the Bogoliubov spectrum. We will numerically demonstrate the occurrence of this instability in Sec.~\ref{sec:numericslandau}; this has not yet been observed experimentally as it requires a very high nonlinearity and hence a very strong light intensity. 
	
	In the intermediate region $0 \leq \Gamma I_0 \leq 0.5$ of stability, the Bogoliubov dispersion is purely real; still, the dispersion is dramatically affected by the interactions. This can be seen by noting that in the linear regime ($\Gamma=0$), straightforward trigonometric algebra shows that the positive-norm $\eta>0$ components of the Bogoliubov dispersion~\eqref{eq:bog} recover (modulo a shift by the Floquet quasi-energy of the unperturbed state $\theta=- \pi/4$) the band dispersion \eqref{eq:linear_dispersion} in the absence of nonlinearities. Hence, in the linear case, all branches have a parabolic dispersion around $k=0,\pi$.
	
\begin{figure}[!]
		\includegraphics[width=1\linewidth]{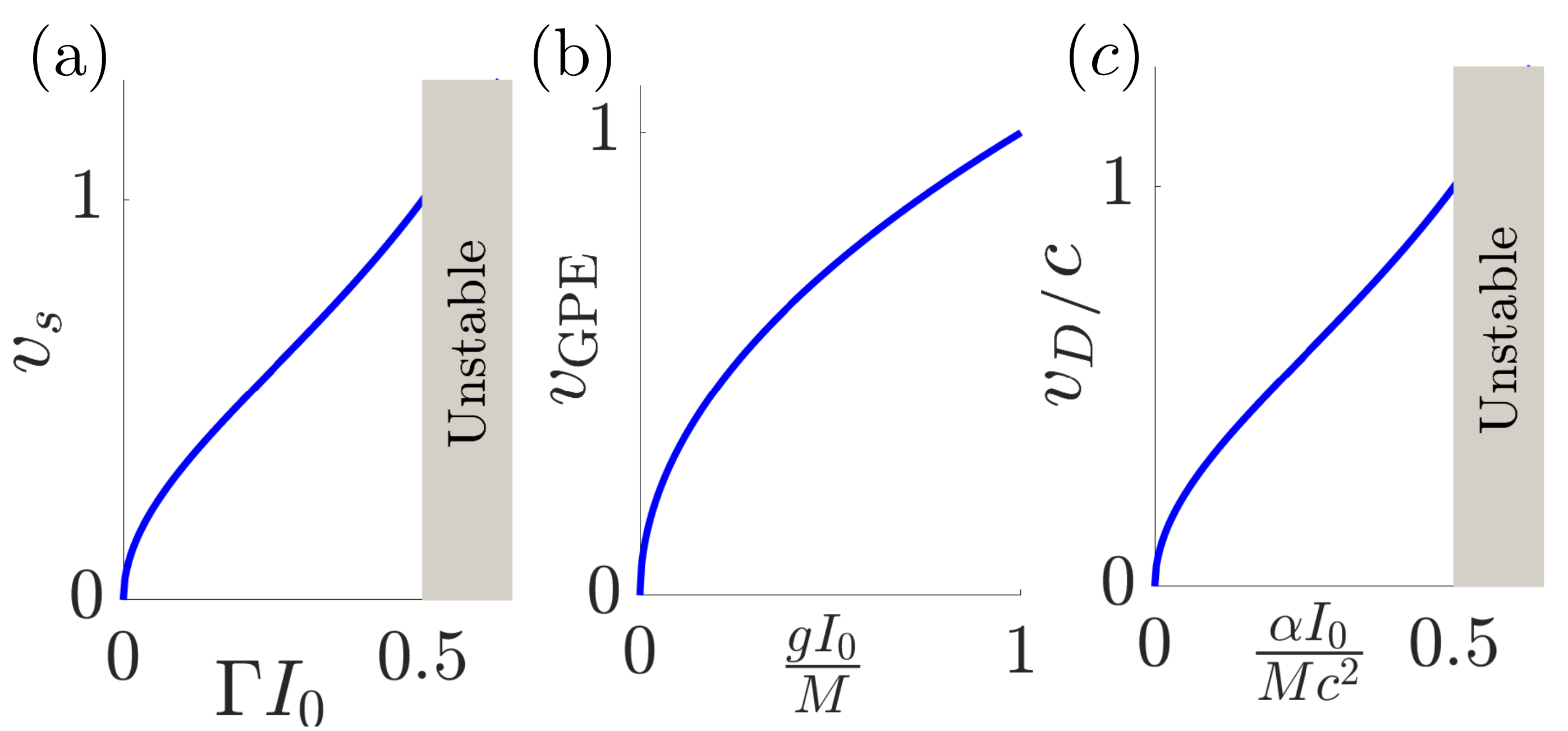}
		\caption{The speed of sound as calculated for (a) the optical mesh lattice, (b) the GPE equation, (c) a Dirac-type relativistic model, inspired by the optical mesh lattice. As can be seen, the speed of sound in the optical mesh lattice and the Dirac-type model share the same functional form, and are limited by the stability of the field, in contrast to the GPE result which increases without bound. }\label{diracsound}	
	\end{figure}

	However, in the nonlinear regime, in the vicinity of $k=0$, the upper (Goldstone) branch (i.e. with $j=-1$, $\eta=1$ in the notation introduce above) acquires a linear form: 
	\begin{eqnarray}
		\theta (k) =-\sqrt{\frac{\Gamma I_0}{1- \Gamma I_0} } |k| + \mathcal{O} (|k|^3) , \label{eq:so}
	\end{eqnarray}
	indicating that the long wavelength excitations of the system are phonon-like, moving with a speed of sound:
	\begin{eqnarray}
		v_{\text{s}} = \sqrt{\frac{\Gamma I_0}{1- \Gamma I_0} }, \label{eq:sound}
	\end{eqnarray}
	as indicated by the blue dashed lines in Fig.~\ref{bogdispersion} and as shown in Fig.~\ref{diracsound}. Interestingly, this speed of sound appears to diverge for $\Gamma I_0 \rightarrow 1$; however, the system is also unstable at such high nonlinearities, and so, in practice, the  speed of sound is bound by its maximum value of $v_{\text{s}} \rightarrow 1$ that is reached for $\Gamma I_0 \rightarrow 0.5$. This is still twice as large as the maximum speed which linear excitations on the empty lattice can acquire.
	As we now discuss, this is unlike the usual behaviour of superfluids, but instead can be viewed as a relativistic effect, stemming from the form of the evolution equations. 
	
	\begin{figure*}[!]
		\includegraphics[width=0.99\linewidth]{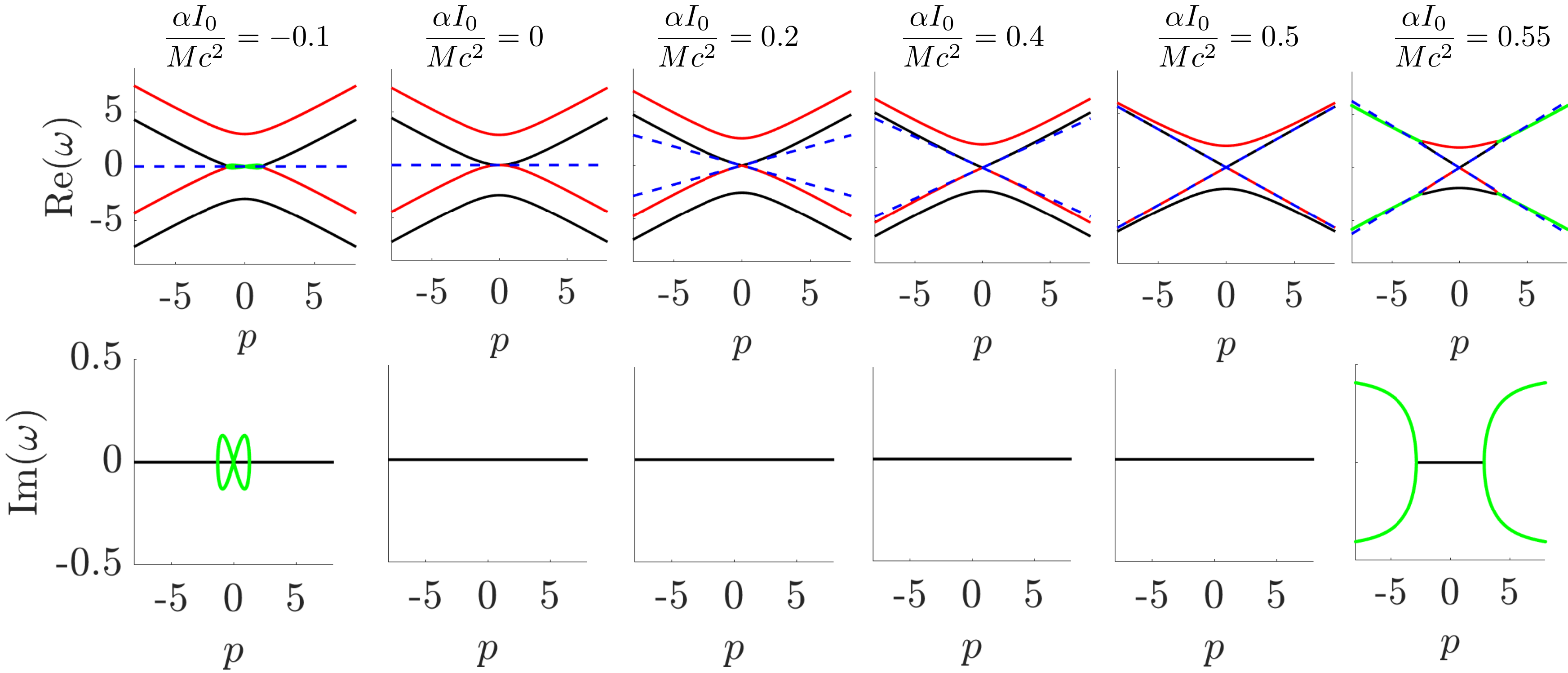}
		\caption{The real ({\it top panels}) and imaginary ({\it bottom panels}) parts of the Bogoliubov dispersion for the relativistic Dirac equation (Eq.~\ref{eq:diraceq}), for different values of the nonlinearity parameter:  $\alpha I_0 / Mc^2 =-0.1, 0, 0.2, 0.4, 0.5, 0.55$ ({\it left to right}). All other parameters $M=-2\sqrt{2}$, $c=1/\sqrt{2}$, $I_0=1$ and $\alpha = - \Gamma \sqrt{2}$ are chosen in analogy with the optical mesh lattice. The phonon-like linear dispersions (Eq.~\ref{eq:diracspeed}) are marked by the blue dashed lines. The black/red/green ({\it in greyscale resp. black/grey/light grey}) color of each curve indicates the negative/positive/zero value of the Bogoliubov norm of the band, respectively. For $\alpha I_0 / Mc^2<0$ and $\alpha I_0 / Mc^2>0.5$, the curves develop a non-zero imaginary part, indicating dynamical instability. As can be seen, the low-momentum behaviour is qualitatively very similar to that of the optical mesh lattice (Fig.~\ref{bogdispersion}), while that at higher momenta deviates as the Dirac equation (Eq.~\ref{eq:diraceq}) describes a continuum model instead of a lattice. }
		\label{dirac}	
	\end{figure*}
	
	\subsection{Comparison with non-relativistic and relativistic quantum fluids}\label{sec:nonlineardirac}
	
	We now compare the above results, firstly, with the usual non-relativistic Gross-Pitaevskii equation (GPE) describing, for example, weakly-interacing cold atomic Bose-Einstein condensates~\cite{pitaevskii2016bose}, and secondly, with a type of Dirac relativistic system, which is motivated from the continuum limit of the linear optical mesh lattice.
	
	The GPE can be written as~\cite{pitaevskii2016bose}
	\begin{eqnarray}
		i \frac{\partial \psi }{\partial t} =  - \frac{1}{2 M} \frac{\partial^2 \psi}{\partial x^2} + g |\psi|^2 \psi  \label{eq:gpe}
	\end{eqnarray}
	where $\psi$ is the wave function, $M$ is the particle mass, $g$ is the mean-field interaction parameter and we have set $\hbar=1$. Expanding with respect to small perturbations to the light field as $\psi = \psi_0 + \delta \psi$, the well-known Bogoliubov dispersion for a homogeneous system is found as~\cite{pitaevskii2016bose,CastinLectures}
	\begin{eqnarray}
		\omega =\pm \sqrt{ \frac{g I_0 }{M} p^2 + \left(\frac{p^2}{2 M}\right)^2}  \label{eq:gpedispersion}
	\end{eqnarray}
	where $I_0=|\psi_0|^2$ is the unperturbed density, and $\omega$ and $p$ are, respectively, the frequency and momentum of the elementary excitations. For small momenta, this dispersion can be expanded as $\omega(p) = v_{\text{GPE}} p$, where the speed of sound is given by
	$$v_{\text{GPE}} = \sqrt{{g I_0 }/{ M}},$$
	as shown in Fig.~\ref{diracsound}. 
	As in the optical mesh lattice, the system is unstable if $g<0$, corresponding to attractive interactions between particles. However, unlike the optical mesh lattice, there is no second instability in this dispersion when $g>0$, and $ v_{\text{GPE}}$ simply increases as $\sqrt{g}$ without either a maximum or divergence at finite nonlinearity strength. 

	To find a system which is more similar to the optical mesh lattice, we return first to the underlying evolution equations~\eqref{eq:nonlinear_1}-\eqref{eq:nonlinear_2}. In the absence of nonlinearities, 
	$\Gamma=0$, these equations can be written, after some algebra and relabelling, as~\cite{martinthesis}: 
	\begin{eqnarray}
		\sqrt{2} (u^{m+1}_n - u_n^{m-1}) &=& (u^m_{n+1} - u^m_{n-1}) + 2 i v^m_{n}, \nonumber \\
		\sqrt{2} (v^{m+1}_n - v_n^{m-1}) &=& -(v^m_{n+1} - v^m_{n-1}) + 2 i u^m_{n},
	\end{eqnarray}
	where discretised derivatives with respect to the time step and position are recognisable on the LHS and RHS respectively.  
	Taking the continuum limit, these equations can be written compactly as
	\begin{eqnarray}
		i \frac{\partial }{ \partial t} 
		\left( \begin{array}{c} \psi \\  \chi \end{array} \right) =  \frac{i \sigma_z}{\sqrt{2}}\frac{\partial }{ \partial x}  \left( \begin{array}{c} \psi \\  \chi \end{array} \right) - \sqrt{2} \sigma_x
		\left( \begin{array}{c} \psi \\  \chi \end{array} \right), \label{eq:diraceq}
	\end{eqnarray} 
	where we have taken $\Delta t\! =\!\Delta x\! =\!1$, and where the two components of the vector $(\psi, \chi)^T$ reflect the short and long loop degrees of freedom, $(u,v)^T$. For compactness, we have introduced the Pauli matrices $\sigma_z$ and $\sigma_x$. From the first-order spatial derivative on the RHS, this can be recognised as a type of relativistic Dirac equation for a two-component spinor wave-function. In the nonlinear regime, such a simple direct mapping is no longer possible. Nonetheless, motivated by this analogy, we consider a nonlinear version of~\eqref{eq:diraceq} as
	\begin{eqnarray}
		i \frac{\partial }{ \partial t} \!
		\left(\!\begin{array}{c} \psi \\  \chi \end{array}\!\right)\! =\! \left[  i c  \sigma_z \frac{\partial }{ \partial x} + M c^2 \sigma_x
		+ \alpha \left(\!\begin{array}{cc} |\psi|^2 &0 \\ 0&  |\chi |^2\end{array}\!\right) \right]\!\left(\!\begin{array}{c} \psi \\  \chi \end{array}\!\right)\!,\qquad
	\end{eqnarray}
	where we have introduced a speed of light $c=1/\sqrt{2}$ and a mass $M=-2\sqrt{2}$. The parameter $\alpha=-\Gamma \sqrt{2}$ quantifies the strength of the nonlinearity, whose form is chosen such that each component interacts with itself but not with the one in the other loop. Note that we did not introduce units for time and space in Eq.~\ref{eq:diraceq} and therefore the quantities introduced in the Dirac equation lack a unit as well. Similar equations have also been studied to describe cold atoms in a honeycomb lattice~\cite{haddad2009nonlinear,haddad2011relativistic}. 
	
	As in Sec.~\ref{sec:band}, we begin by checking the linear ($\alpha=0$) dispersion
	by using a plane-wave ansatz:
	\begin{eqnarray}
		\left( \begin{array} {c} \psi (x,t) \\ \chi (x,t) \end{array} \right)= \left( \begin{array} {c} \psi_0 \\\chi_0 \end{array} \right) e^{i (Px - \Omega t)} 
	\end{eqnarray}
	where $P$ is the corresponding momentum and $\Omega$ is the frequency. This leads to $\Omega^2 = P^2 c^2 + M^2 c^4$, as expected for a relativistic Dirac model. As this is a continuum rather than a lattice model, none of the sub-lattice issues discussed in detail in Sec.~\ref{sec:band} are any longer relevant.
	
	The Bogoliubov dispersion of collective excitations can then be calculated, as before, by considering small perturbation on top of a stationary and initially unperturbed field, i.e. such that $\psi \rightarrow \psi_0 + \delta \psi $ and  $\chi \rightarrow \chi_0 + \delta \chi $. Similar to Sec.~\ref{sec:bog}, we assume the unperturbed light field to have a finite power and to be at rest at $P=0$, as described by:
	\begin{eqnarray}
		\left( \begin{array}{c} \psi (x,t) \\ \chi (x,t)  \end{array} \right)
		=\left( \begin{array}{c} \sqrt{I_0} \\ \mp \sqrt{I_0} \end{array} \right) e^{\pm i M c^2 t} e^{- i \alpha I_0 t},
	\end{eqnarray}
	where the signs correspond to states with $\Omega (P\!=\!0)\!=\! \mp M c^2 + \alpha I_0$ respectively. Note that in terms of the mesh lattice parameters introduced above, this corresponds to having $\Omega (P\!=\!0)\!=\! \pm \sqrt{2}  -  \sqrt{2} \Gamma I_0 $.
	
	As in the optical mesh lattice, our next step is to consider weak perturbations on top of a uniform field at rest in the lower band, i.e, at $\Omega (P\!=\!0)\!=\! - \sqrt{2}  -  \sqrt{2} \Gamma I_0 $. The small perturbations, $\delta \psi (x, t)$ and $\delta \chi (x, t)$, are described by a temporally and spatially periodic ansatz:
	\begin{eqnarray}
		\left( \begin{array}{c} \delta \psi (x, t) \\  \delta \chi (x, t)\end{array} \right)
		=\left( \begin{array}{c} A_\psi e^{-i (\omega t - p x)} + B_\psi^* e^{i (\omega t - p x)} \\ A_\chi e^{-i (\omega t - p x)} + B_\chi^* e^{i (\omega t - p x)} \end{array} \right)
	\end{eqnarray}
	where $A_\psi,A_\chi,B_\psi,B_\chi$ are constant coefficients and where $\omega$ and $p$ are the frequency and momentum of the Bogoliubov excitations. Inserting this ansatz into the nonlinear Dirac equation~\eqref{eq:diraceq} and linearizing in the small perturbations, one gets the following matrix equation for the collective eigenmodes,
	\begin{equation}
		\omega \vec{\Phi}_D = \mathcal{L}_p \vec{\Phi}_D
	\end{equation}
	with $\vec{\Phi}_D=(A_\psi,A_\chi,B_\psi,B_\chi)^T$ and the Bogoliubov matrix
	\begin{widetext}
		\begin{equation}
			\mathcal{L}_p = \left( \begin{array}{cccc} 
				\alpha I_0 + c p - c^2 M &  c^2 M  & \alpha I_0  & 0 \\ 
				c^2 M  &  \alpha I_0 - c p - c^2 M & 0 & \alpha I_0    \\
				-\alpha I_0  & 0 & -\alpha I_0 + c p + c^2 M & -c^2 M       \\
				0 &  -\alpha I_0 & -c^2 M  & -\alpha I_0 - c p + c^2 M     
			\end{array} \right).\nonumber
		\end{equation}
	\end{widetext}
	Solving this equation leads to the Bogoliubov dispersion relation: 
	\begin{eqnarray}
		&&\omega^2/c^2 = 2 M^2 c^2 - 2 M \alpha I_0 + p^2 \nonumber \\ &&\pm 2 \sqrt{M^4 c^4 - 2 M^3 \alpha I_0 c^2 +  (M \alpha I_0)^2 + M p^2 (M c^2 - 2  \alpha I_0 )}. \nonumber \end{eqnarray}
	As in the nonlinear optical mesh lattice [c.f. Fig.~\ref{bogdispersion}], this has two positive and two negative branches, which are plotted in Fig.~\ref{dirac}. These branches are plotted in black/red/green depending on the negative/positive/zero value of the Bogoliubov norm $|A_\psi|^2+|A_\chi|^2-|B_\psi|^2-|B_
	\chi|^2$. As before, branches are organized in pairs with opposite momentum $p$ and frequency $\omega$ and opposite norm, and one pair of these branches goes through the $p=0$ and $\omega=0$ point to satisfy Goldstone's theorem. As in the nonlinear optical mesh lattice, the Bogoliubov dispersion becomes unstable when $\alpha I_0 / M c^2 <0$ or $\alpha I_0 > Mc^2 /2$: the former instability stems from the focusing nonlinearity, while the latter one arises from a level attraction between pairs of opposite-norm Bogoliubov branches. 
	
	The speed of sound for low momenta excitations is then
	\begin{eqnarray}
		v_D = c \sqrt{ \frac{\alpha I_0 }{M c^2 - \alpha I_0 }} , \label{eq:diracspeed}
	\end{eqnarray}
	as also indicated in Fig.~\ref{dirac} and plotted in Fig.~\ref{diracsound}. This is the same functional form as that of the speed of sound for the nonlinear optical mesh lattice~\eqref{eq:sound}. As in that system, the speed of sound diverges when $ \alpha I_0 \rightarrow Mc^2$, but this divergence is cut off by the instability at $\alpha I_0 > Mc^2 /2$, and so instead there is a maximum speed of sound set by the speed of light: $v_{\text{D}} \rightarrow c$ as $\alpha I_0 \rightarrow Mc^2 /2$, as shown in Fig.~\ref{diracsound}. 
	
	In summary, we have found that the Bogoliubov dispersion for the Dirac model and the nonlinear optical mesh lattice are qualitatively similar in all major respects at low momentum. This means that the optical mesh lattice can provide an optical set-up to observe certain relativistic superfluid effects. However, differences naturally appear at high momentum as the Dirac model describes a continuum system while the optical mesh lattice is discrete; as a result, for the former, the frequency and momentum can increase without bound, while for the latter, the dispersion must be periodic (with the subtleties discussed in Sec.~\ref{sec:bog}). 
	
	\section{Measuring the speed of sound} \label{sec:measure}
	
	In the previous Section we laid down the general theory of sound propagation in fluids of light in optical mesh lattices and we have characterized the different regimes. 
	In this Section, we discuss how the above predictions for the speed of sound can be experimentally verified by switching on and off a defect at rest, and observing how the resulting perturbations propagate in the fluid. We begin with numerical simulations for the idealised case of a temporally long and spatially wide defect in a uniform light field. We then bring this closer to experiment by investigating more realistic defect profiles, and asking what happens when such defects move in an expanding rather than a uniform light field. In so doing, we demonstrate that the underlying expansion speed has to be taken into account in order to accurately measure the local speed of sound. This provides provides a theoretical framework and further numerical support for the data analysis protocols that we applied in the main text and the supplemental material of our work~\cite{experiment}. In the analysis of the experimental data, we were forced to focus on qualitative signatures in measurements of the speed of sound due to experimental uncertainties, such as in the overall value of the experimental nonlinearity strength. In the current work, instead, we provide a comprehensive theoretical framework that sets the stage for future more quantitative experimental measurements of the speed of sound.
	
	To create a defect in the optical field, we use the time-dependent phase-shifts in~\eqref{eq:nonlinear_1}-\eqref{eq:nonlinear_2} to imprint a Gaussian defect phase shift in both loops, according to
	\begin{eqnarray}
		\varphi_{n}^m = \phi_{n}^m = \frac{\varphi_0}{\sqrt{2\pi \sigma_n^2}} e^{-(n - n_d)^2 / 2 \sigma_n^2} e^{-(m - m_d)^2 / 2 \sigma_m^2} \label{eq:defectform}
	\end{eqnarray}
	where $\varphi_0$ is the defect amplitude, $\sigma_n$ ($\sigma_m$) is the standard deviation of the defect profile with respect to position (time) and $n_d$ ($m_d$) is the location of the defect peak amplitude in position (time). Throughout this section, we concentrate on the simplest case of a defect that is stationary with respect to the lattice, i.e. $n_d$ is a constant at all time steps, hereafter taken to be $n_d=0$ unless otherwise specified; the case of a moving defect will instead be studied in Sec.~\ref{sec:landau}. 
	
	\subsection{Idealised defects in uniform light fields} \label{sec:sosnumerical}
	
	\begin{figure}[!]
		\includegraphics[width=0.95\linewidth]{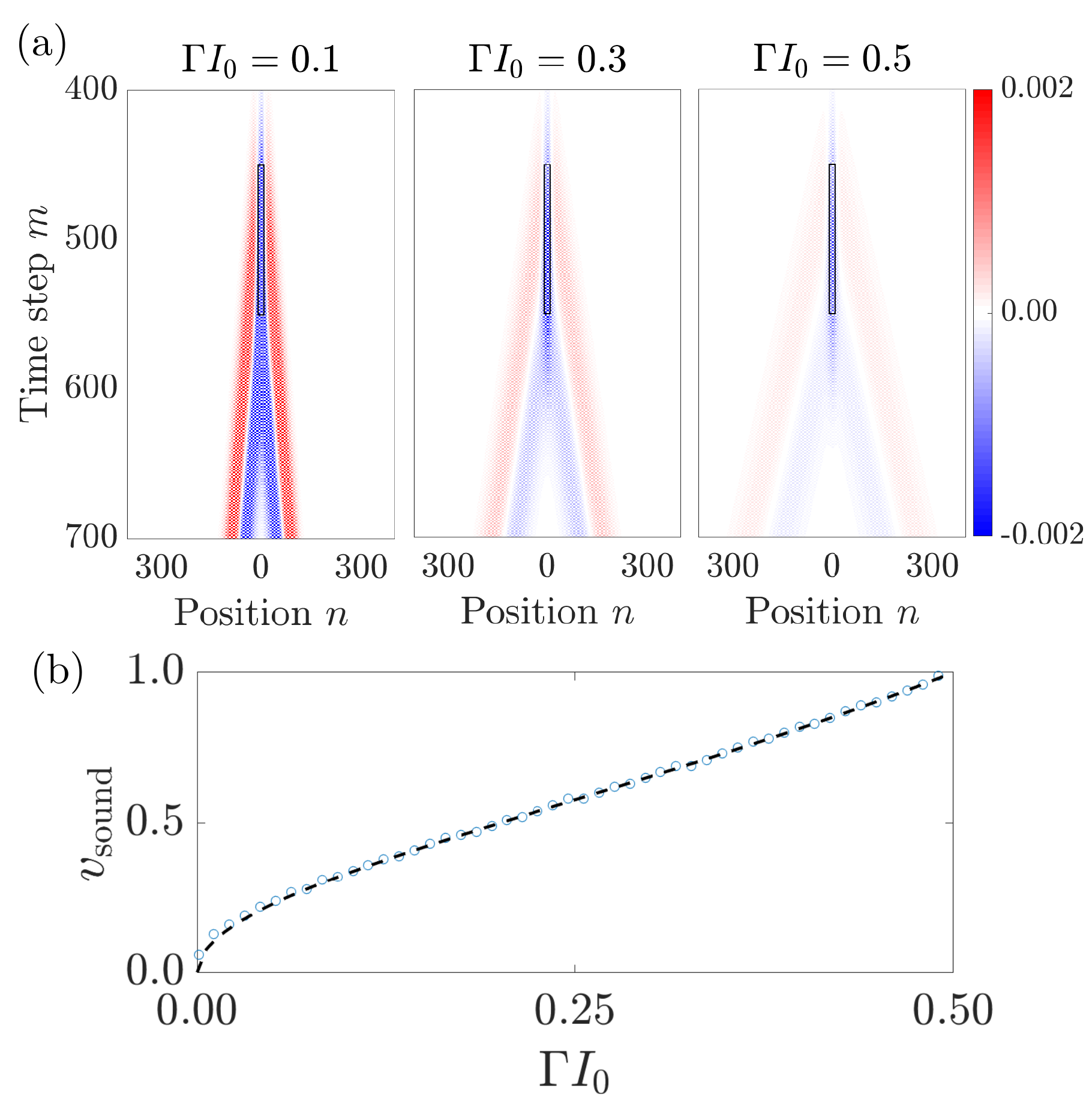}
		\caption{ (a) Spatio-temporal colorplot of the numerical differential intensity (Eq.~\ref{eq:diff}) for a smoothly-varying, weak defect at rest in a uniform light field, with different values of the nonlinearity parameter  $\Gamma I_0 = 0.1, 0.3, 0.5$. The defect profile (Eq.~\ref{eq:defectform}) is indicated by a solid black line, marking one standard deviation, $\sigma_n\!=\!10$ and $\sigma_m\!=\!50$, from the defect peak amplitude of $\varphi_0=0.01$ at $m_d=500$ and $n_d=0$. This smooth defect profile predominantly excites long-wavelength sound waves. (b) The numerical sound velocity \textit{(blue circles)} as a function of the nonlinearity, extracted by tracking the propagation of sound waves in (a) via Eq.~\ref{eq:numericalsound} with $m_1=800$ and $m_2=1000$. This shows excellent agreement with the analytical speed of sound Eq.~\ref{eq:sound}(dashed black line). 
		}\label{theory_sound}	
	\end{figure}
	
	To verify the expected behaviour of the speed of sound, we begin by numerically studying the idealised case, where the defect profile~\eqref{eq:defectform} is very wide and smoothly varying compared to the discrete position and time step, i.e. $\sigma_n, \sigma_m \gg 1$. In this regime, all but the lowest-energy excitations are suppressed, leading to an emission dominated by long-wavelength sound waves. 
	
	To further simplify the data analysis, we also consider the simplest initial condition of a spatially-uniform stationary optical light field, with the form: 
	\begin{eqnarray}
		u_n^0 = v_n^0= \sqrt{I_0}\, \rm{mod} (n,2) , \label{eq:initial}
	\end{eqnarray}
	corresponding to exciting the lower-band eigenstate at $Q=0$. Here,  $I_0$ is the initial intensity in each loop, and the factor of $\rm{mod} (n+m,2)$ (with $m=0$)  highlights the connectivity of the optical mesh lattice,  as physically only either even or odd lattice sites are occupied at any given time step (see Figs.~\ref{overview} and \ref{fig:methods}). 
	
	\begin{figure}[!]
		\includegraphics[width=0.95\linewidth]{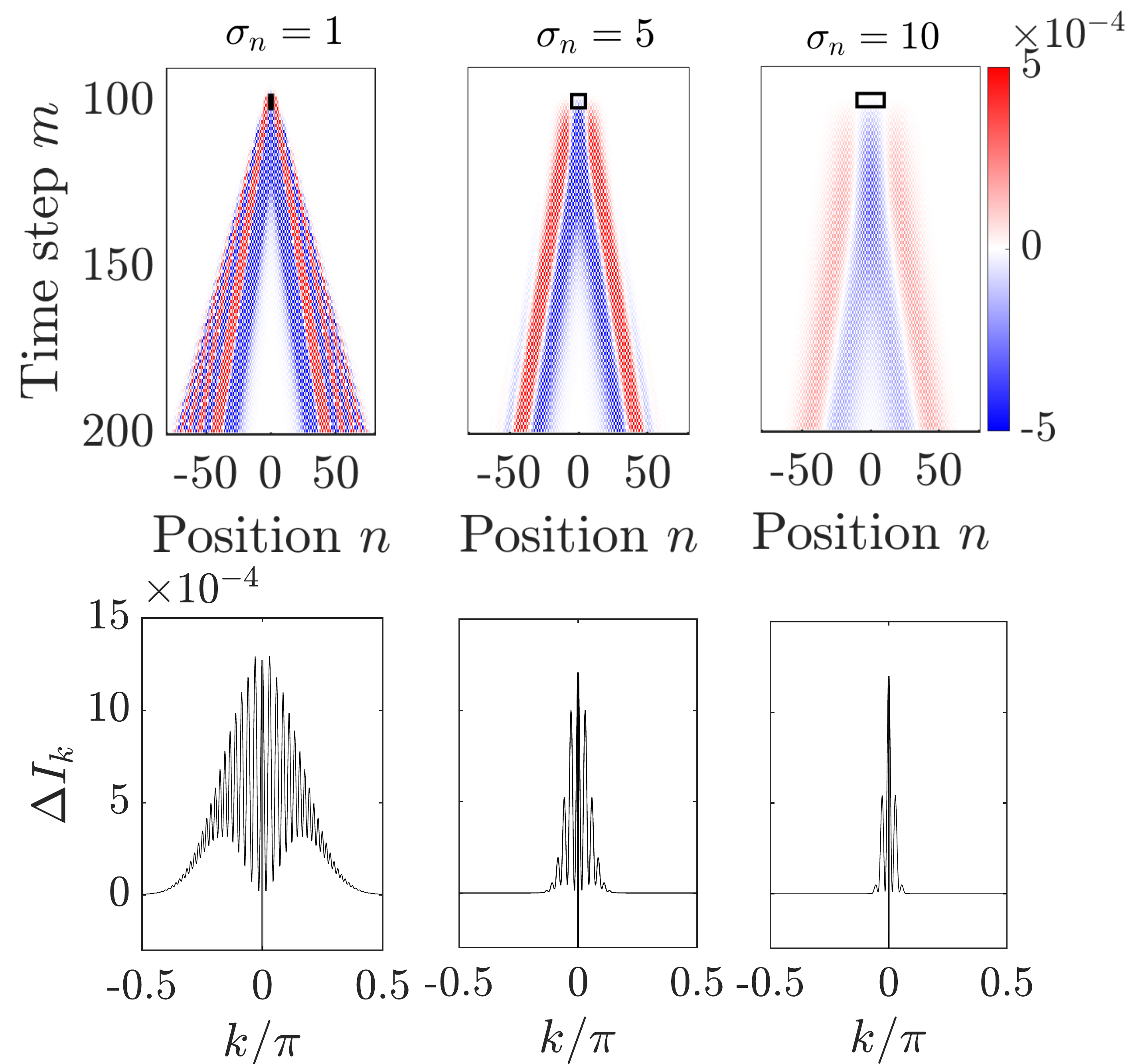}
		\caption{ ({\it Top Row}) Spatio-temporal colorplot of the numerical differential intensity (Eq.~\ref{eq:diff}) for a rapidly-varying, weak defect at rest in a uniform light field, with an increasing spatial width $\sigma_n$ for a nonlinearity $\Gamma I_0 =0.1$ . The defect profile (Eq.~\ref{eq:defectform}) is indicated by a solid black line, marking one standard deviation, $\sigma_n$ and $\sigma_m\!=\!2$, from the defect peak amplitude of $\varphi_0=0.01$ at $m_d=100$ and $n_d=0$. For wide defects, the emission is dominated by low-momentum sound waves, while for narrow defects, there is significant emission of higher-momentum excitations beyond sound-waves. ({\it Bottom Row}) Corresponding plots of the momentum-space differential intensity spectrum $\Delta I_k=\Delta I_k^{m_\text{max}}$ (Eq.~\ref{eq:iq}) numerically calculated at $m_\text{max}=200$ showing that as the defect width decreases, the range of significant Fourier components increases, indicating the importance of excitations at higher momenta. Note that the scale of the y-axis is chosen to highlight the spread of Fourier components but it artificially cuts off the dip in the spectrum at $k=0$.} \label{narrowstat}	
	\end{figure}
	
	The propagation of excitations can be visualised through the normalized differential intensity:
	\begin{eqnarray}
		\Delta I= \frac{I_{{\text{pert.}}} -I_{{\text{unpert.}}}}{2 I_0} \label{eq:diff}
	\end{eqnarray}
	where $I_{{\text{pert.}}}=|u_n^m|^2+|v_n^m|^2$ (or correspondingly $I_{{\text{unpert.}}}$) is the local intensity when a defect is present (absent). The numerical differential intensity, $\Delta I$, is plotted in Fig.~\ref{theory_sound}(a) as a function of $m,n$ for three values of the nonlinearity parameter $\Gamma I_0 =0.1, 0.3, 0.5$. A small defect amplitude, $\varphi_0=0.01$, and large defect widths, $\sigma_n\!=\!10$ and $\sigma_m\!=\!50$, have been chosen so as to create a wide, weak and slowly-varying intensity modulation in the uniform light field. As can be seen, turning on the defect leads to the symmetric emission of propagating intensity peaks from the defect position, while turning off the defect generates propagating intensity dips; of course, the order of the peaks and dips can be switched by changing the sign of the defect amplitude.
	
	As the nonlinearity parameter, $\Gamma I_0$, is increased, we observe two important and related effects in the emission pattern shown in Fig~\ref{theory_sound}(a). Firstly, the speed of sound (Eq.~\ref{eq:sound}) increases with the nonlinearity, as evident from the increasing angle at which the sound waves propagate in these plots of time versus position. Secondly, the magnitude of the differential intensity decreases, indicating that the excitation of sound waves is suppressed at stronger nonlinearities for a stationary defect.

	\begin{figure*}[!]
		\includegraphics[width=0.9\linewidth]{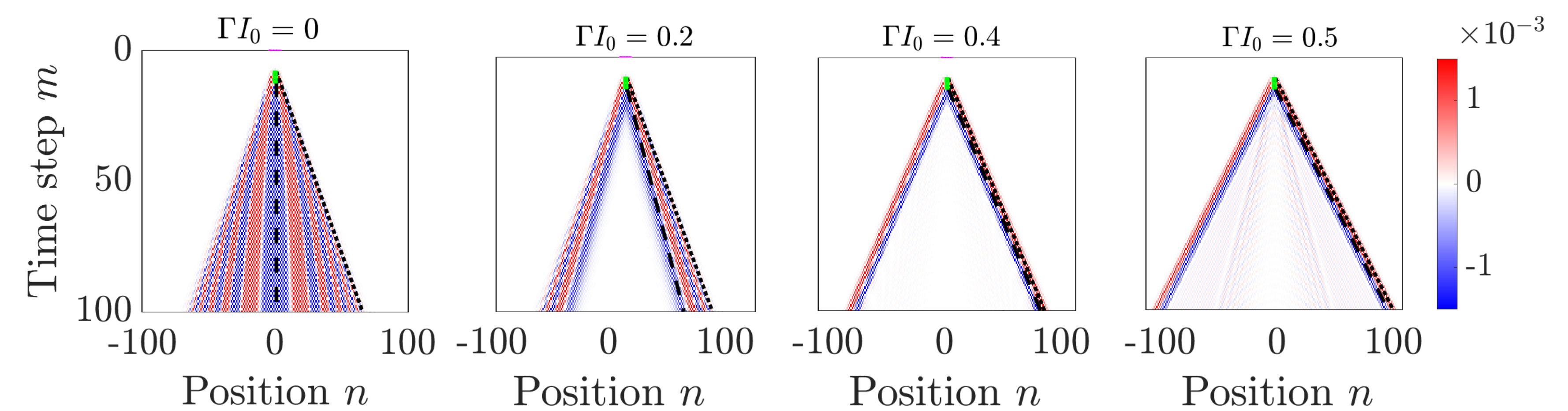}
		\caption{Spatio-temporal colorplot of the differential intensity (Eq.~\ref{eq:diff}) for a rapidly-varying, weak defect at rest in a uniform light field, for different values of the nonlinearity parameter $\Gamma I_0 =0, 0.2, 0.4, 0.5$. The defect profile (Eq.~\ref{eq:defectform}) is indicated by a solid green line, marking one standard deviation, $\sigma_n\!=\!1$ and $\sigma_m\!=\!2$, from the defect peak amplitude of $\varphi_0=0.01$ at $m_d=10$ and $n_d=0$.  The analytical speed of sound (Eq.~\ref{eq:sound}) and the maximum group velocity (Eq.~\ref{eq:vg} and see Fig.~\ref{groupvelocity}) are marked with dashed and dotted black lines emanating from the bottom-right and top-right corners of the defect profile, respectively. At low nonlinearities, the emission is effectively bounded between the dashed and dotted lines, while at high nonlinearities, the emission is dominated by sound waves, which move with the maximum group velocity of all Bogoliubov excitations. }\label{cones}
	\end{figure*}
	
	To calculate the speed of sound in the long-wavelength regime, we can simply track the propagation of the emitted intensity peaks (or dips) to estimate: 
	\begin{eqnarray}
		v_{\rm{sound}}^{\rm{num.}} = \frac{\Delta n}{\Delta m} = \frac{n_{\rm{max}}(m_2)-n_{\rm{max}}(m_1)}{m_2-m_1} \label{eq:numericalsound}
	\end{eqnarray}
	where $\Delta n=  n_{\rm{max}}(m_2)-n_{\rm{max}}(m_1)$ is the peak displacement over the time step interval $\Delta m= m_{2}-m_{1}$. This is plotted in Fig~\ref{theory_sound}(b), with $m_1=800$, $m_2=1000$, corresponding to late times after the defect has been turned off. This numerically-estimated speed of sound shows excellent agreement with the analytical expression from Bogoliubov theory (Eq.~\ref{eq:sound}).

	\subsection{Realistic defects in uniform light fields} \label{sec:real}
	\begin{figure*}[!]
		\includegraphics[width=1\linewidth]{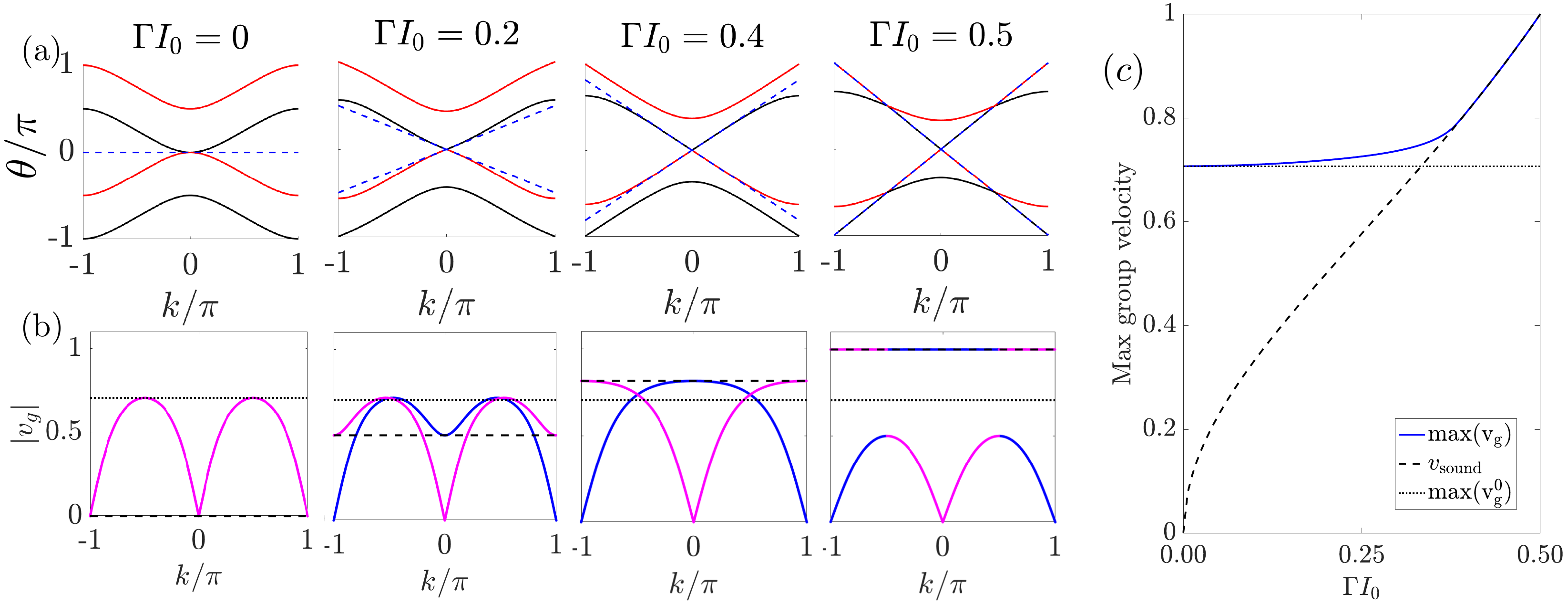}
		\caption{(a) The Bogoliubov dispersion for the optical mesh lattice (Eq.~\ref{eq:bog}), for the different values of the nonlinearity parameter $\Gamma I_0 =0, 0.2, 0.4, 0.5$ considered in Fig.~\ref{cones}, as previously shown in Fig.~\ref{bogdispersion}. The blue dashed lines indicate the phonon-like linear dispersions with the speed of sound given by Eq.(\ref{eq:so}). (b) The absolute value of the group velocity, $|v_g| = | \partial \theta / \partial k |$, for the same values of the nonlinearity used in (a), shown with a solid blue (dark grey in greyscale) line for the two middle Bogoliubov bands (i.e. those meeting at $\theta=0$ in (a)) and shown with a solid purple (light grey in greyscale) line for the other two outer bands. Note that in the linear regime, for $\Gamma I_0=0$, the blue and purple lines are identical. In each plot, the analytical speed of sound (Eq.~\ref{eq:sound}) and the maximum group velocity in the linear regime ($\rm{max}(v_g^0)= 1/\sqrt{2}$) are marked with, respectively, dashed and dotted black lines. (c) Plot of the maximum group velocity extracted from (b) as a function of the nonlinearity parameter $\Gamma I_0$ (solid blue line). The black lines indicate the speed of sound (dashed) and the maximum group velocity in the linear regime (dotted).  
		}\label{groupvelocity}	
	\end{figure*}

	While the above numerical results are encouraging, the parameters required to reach the long-wavelength regime are far beyond current experiments where propagation times are typically limited to a few hundred time steps and the initial optical field has a width only on the order of ten positions.
	As a result, experimentally-realistic defects with $\sigma_n, \sigma_m \simeq 1$ will significantly excite both sound waves and higher-momenta Bogoliubov excitations. However, as we now show numerically, it is still possible to extract an accurate measurement of the speed of sound by tracking the propagation of excitations with the lowest group velocity.  
	
	For realistic defects, the oscillatory emission patterns can be much richer than for idealised defects, as shown, for example, in Fig.~\ref{narrowstat}, where we progressively increase the spatial width of a rapidly-varying defect in a uniform light field. As can be seen, for wide defects (right), the emission is dominated by low-momentum sound waves, similar to Fig.~\ref{theory_sound}, while for narrow defects (left panel), there are many more peaks and dips, moving with velocities greater than the speed of sound. To understand this, we define the differential intensity spectrum according to: 
	\begin{eqnarray}
		\Delta I^m_k \equiv { (|u^m_k|^2 + |v^m_k |^2)_{\rm{pert.}} - (|u^m_k|^2 + |v^m_k|^2)_{\rm{unpert.}}}
		\qquad \label{eq:iq}
	\end{eqnarray}
	where $u^m_k$ and $v^m_k$ are the Fourier transforms with respect to the 1D synthetic lattice spanned by $n$ of the short- and long-loop amplitudes (respectively $u^m_n$ and $v^m_n$) at a given time step $m$. This is plotted in the bottom row of Fig.~\ref{narrowstat} for different defect widths. As can be seen, decreasing the defect width in space, increases the range of Fourier components excited, leading to a significant population of higher-momentum states. These perturbations with a much higher spatial frequency also propagate with a higher velocity than the long wavelength sound waves (see Fig. 4). Note that the spectrum is plotted here for $-\pi/2 \leq k < \pi/2$ [see discussion in Sec.~\ref{sec:bog}].
	
	The dramatic nonlinearity-dependence of the spatial shape of the excitation pattern emitted by a realistic defect is illustrated in Fig.~\ref{cones}.
	In the linear regime (left panel), the excitation pattern consists in a fan of fringes extending up to an outer ``light-cone", determined by the maximum group velocity of the linear waves  (dotted black line),
	\begin{equation}
		{\rm{max}}(v_g^0)={\rm{max}}\left(\frac{\partial\vartheta}{\partial Q}\right)=\frac{1}{\sqrt{2}}
	\end{equation}
	For low nonlinearities (center-left panel), the excitation pattern is bounded by an inner ``sound-cone" within which excitations are significantly suppressed (dashed black line) and whose position is set by the minimum group velocity of the low-momentum Bogoliubov modes around $\theta\!\approx\!0$, corresponding to the speed of sound waves~\eqref{eq:sound}. The visible pattern then extends outwards from this inner limit because of the higher group velocity of some higher momentum modes. From the outside, the pattern for low nonlinearities is effectively limited by an outer ``light-cone" (dotted black line), determined by the maximum group velocity:
	\begin{eqnarray} 
		{\rm{max}}(v_g) ={\rm{max}}\left( \frac{\partial \theta}{\partial k}\right) \label{eq:vg}
	\end{eqnarray}
	of the Bogoliubov dispersion. Note that due to the non-vanishing width and duration of the defect, excitations are not emitted from a single point in time and space. To take this into account, we plot the inner ``sound-cone" in Fig.~\ref{cones} starting from the bottom-right corner of the defect profile, as this approximately tracks emission from the defect at late times, and the outer ``light-cone" starting from the top-right corner of the defect, as this represents emission  at earlier times. 
	
	The dependence of both the sound velocity and  the maximum group velocity on the nonlinearity is shown in the right panel of Fig.~\ref{groupvelocity}.  On the one hand, increasing the nonlinearity increases the speed of sound, shifting the inner ``sound-cone" outwards. On the other hand, at low nonlinearities, the maximum group velocity of the Bogoliubov modes remains well-approximated by its value in the linear regime, $\rm{max}(v_g^0 )=1/\sqrt{2}$, such that that the outer ``light-cone" does not significantly change, and so the oscillating emission pattern is confined within a narrower region. 
	
	At high enough nonlinearities ($\Gamma I_0 \geq 1/3$, center-right and right panels in Fig.\ref{cones}), the speed of sound exceeds the maximum value of the group velocity in the linear regime as shown in Fig.~\ref{groupvelocity}, and the inner and outer bounds merge; in this limit, sound waves dominate, and the emission pattern resembles the one of long and wide defects shown in Fig.~\ref{theory_sound}. Interestingly, the defect could also excite slowly-propagating modes either from the upper Bogoliubov band or from around $k=\pm \pi$ (see Fig.~\ref{bogdispersion}), which would appear within the inner sound-cone; however, the amplitudes of excitations within this region are hardly visible for the defect parameters considered in Fig.~\ref{cones}.
	
	The more complicated oscillatory emission pattern for realistic defect profiles shown in Fig.~\ref{cones} requires a more careful approach to quantitatively extract the speed of sound, as compared to the simple estimate provided by Eq.~\ref{eq:numericalsound}. Guided by the identification of the inner ``sound cone" in this regime, we ask whether sound waves can be identified with the innermost large and smooth feature of the emission pattern. To ascertain  this, we extract the speed of these features, and show that this indeed exactly matches the analytical speed of sound.

	\begin{figure}[h]
		\includegraphics[width=0.9\linewidth]{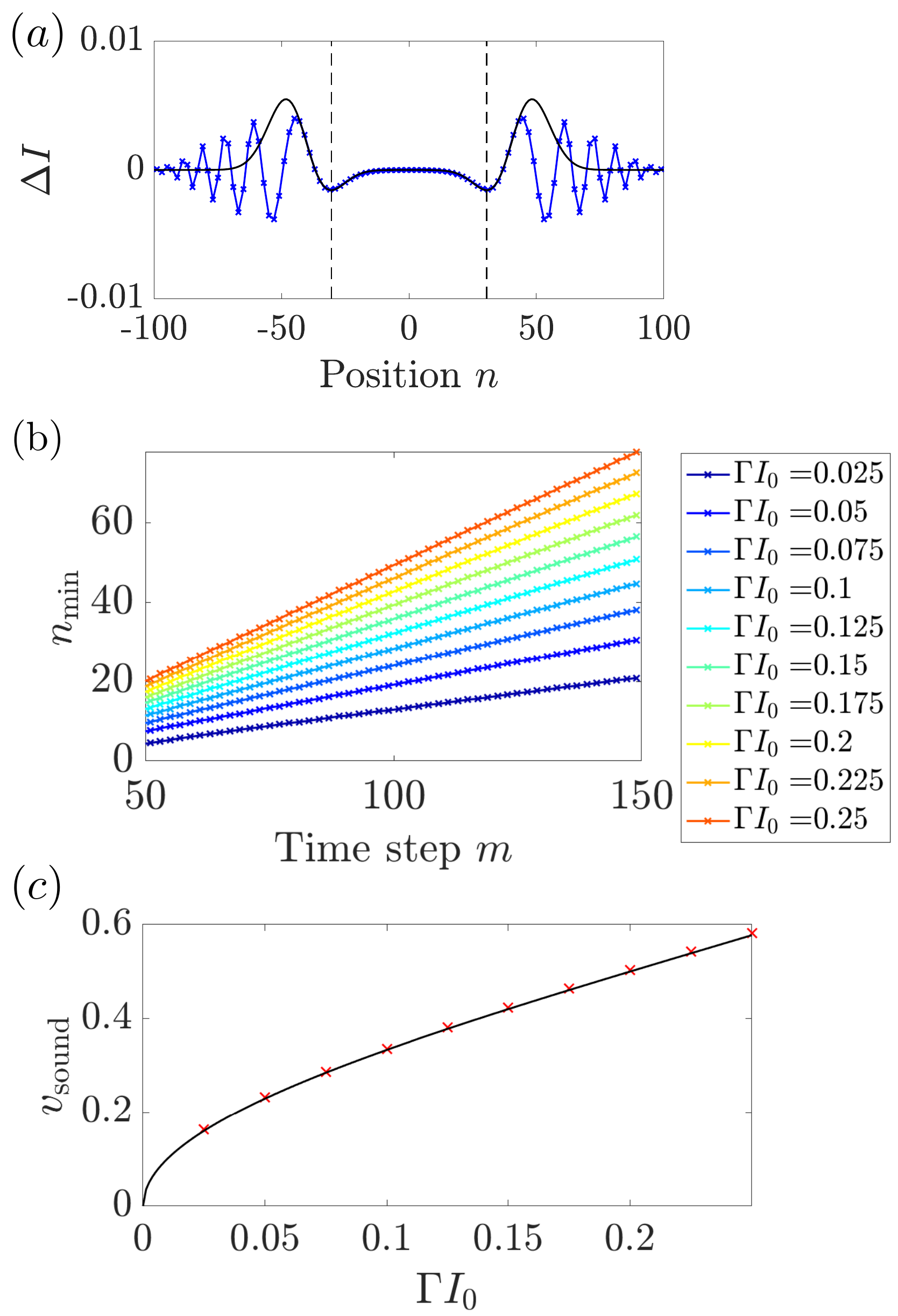}
		\caption{(a) Example cut of the normalised differential intensity at the final time step $m=150$ for $\Gamma I_0 =0.05$. Blue points represent the numerical data and black line is the fitting function~\eqref{eq:fit}. Vertical dashed lines indicate the symmetric minima positions as extracted from the fitting function. The defect is chosen to have spatial and temporal widths of $\sigma_n\!=\!2$ and $\sigma_m\!=\!2$ and a peak amplitude of $\varphi_0=0.01$, and to be centered at $m_d=11$ and $n_d=0$. (b) The extracted absolute positions of the innermost large minima over time from the fits as in panel (a) for a range of different nonlinearities. (c) The numerical estimate for the speed of the sound (red crosses) as a function of the effective initial intensity as calculated from panel (b) by applying a linear fitting function~\eqref{eq:lin} from $m=65$ onwards. This is in excellent agreement with the analytical speed of sound~\eqref{eq:sound}, plotted with a solid black line. 
		}\label{nonadiabatic}	
	\end{figure}

	To measure the speed of the innermost large perturbations, we fit the central region of the numerical data at each time step with the ansatz: 
	\begin{eqnarray}
		f_1(n) &=&( a_1 + b_1(n-c_1))e^{-(n-c_1)^2/d_1^2} \nonumber \\
		&&+ ( a_1 - b_1(n+c_1))e^{-(n+c_1)^2/d_1^2} \label{eq:fit}
	\end{eqnarray}
	where $a_1, b_1, c_1, d_1$ are fitting parameters~\cite{experiment}. This function is chosen as it can reproduce the largest features of the center of the perturbation pattern: namely two inner-most minima (maxima) followed by two maxima (minima) for a positive (negative) defect amplitude. An example fit is shown in Fig.~\ref{nonadiabatic}(a), where the fitting region extends between the two innermost maxima. Using such fits, we numerically extract the positions of the inner-most large minima, as shown in Fig.~\ref{nonadiabatic}(b). As can be seen, the position steadily increases over time corresponding to the excitations propagating outwards from the center. To then measure the speed, we fit the minima positions with the linear function:
	\begin{eqnarray}
		f_2 (m) = a_2 m + b_2 \label{eq:lin}
	\end{eqnarray}
	where $a_2$ and $b_2$ are fitting parameters. We can then associate the value of $a_2$ with the speed of sound as is plotted in Fig.~\ref{nonadiabatic}(c). As can be seen, these numerical results are in excellent agreement with the analytical speed of sound~\eqref{eq:sound}, suggesting that we can indeed identify these large minima with sound waves.

	\subsection{Realistic defects in expanding light fields} \label{sec:expan}
	
\begin{figure}[!]
		\includegraphics[width=1\linewidth]{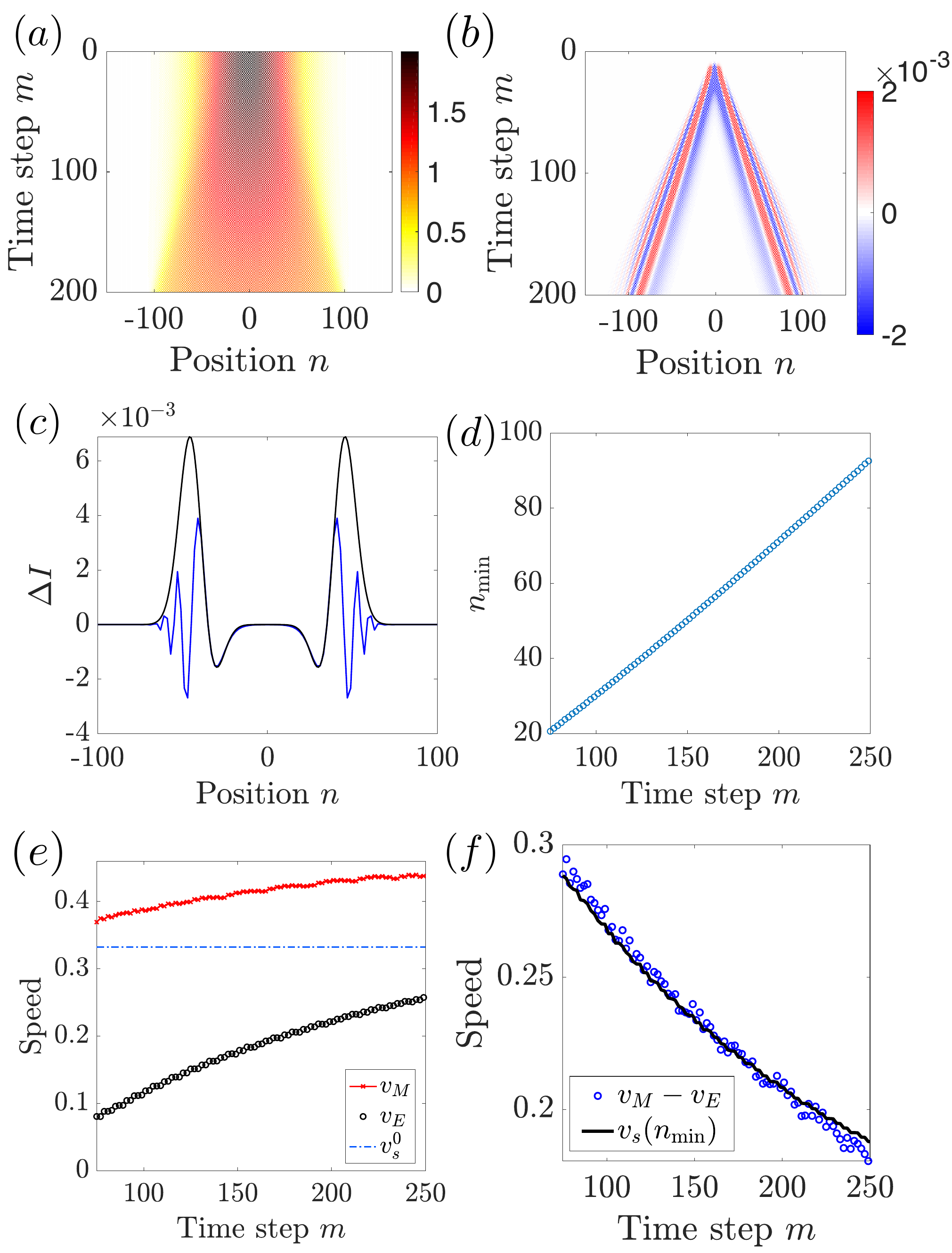}
		\caption{(a) Expansion of a Gaussian optical field over time, with $\Gamma I_0 =0.1$ and $\sigma_G =50$. (b) Spatio-temporal colorplot of the numerical differential intensity after applying a realistic narrow defect, with $\sigma_n\!=\!2$, $\sigma_m\!=\!2$, $\varphi_0=0.01$, $m_d=11$ and $n_d=0$. (c) Fit of a cut of (b) at $m=100$: the blue points are the numerical data and the black curve is the fitting function~\eqref{eq:fit}. (d) Position of the innermost minimum of the fitting function at each time step. (e) Plot of the numerically-estimated drag speed $v_E$ ({\it black dots}), due to the expansion of the condensate, and of  the estimated instantaneous speed of the intensity minimum, $v_M$ ({\it red dots}) as obtained from a moving linear fit~\eqref{eq:lin}, taken over eleven time steps centred around $m$. There are big discrepancies between $v_M$ and the expected analytical value $v_s^0$ of the speed of sound for the peak of the initial intensity ({\it blue dashed line}). (f) Estimate $v_s = v_M - v_E$ for the speed of sound  ({\it blue dots}), which is now in excellent agreement with the analytical speed of sound $v_s (n_{\text{min}})$ using the local unperturbed intensity at the position of the intensity minimum  ({\it solid black line}).}\label{expan1}	
	\end{figure}

	A further complication comes from the fact that in the experiments~\cite{experiment} the initial light field is not spatially uniform, as we have assumed above, but is prepared with some localised profile. In the absence of any other potentials, the light field will therefore expand over time. This expansion is shown, for example, in Fig.~\ref{expan1}(a) for $\Gamma I_0 =0.1$, starting from an initially Gaussian profile:
	\begin{eqnarray}
		u_n^0 \!= \!v_n^0\!=\! \sqrt{I_0} \rm{mod} (n,2) e^{- n ^2 / 2 \sigma_G^2}
	\end{eqnarray}
	where $\sigma_G =50$ is the Gaussian width and the factor of $\rm{mod} (n,2) $ again accounts for the diamond connectivity of the optical mesh lattice. As compared to the Light Walk shown in Fig.~\ref{overview}(b,d), it displays an overall Gaussian intensity profile and a spatially constant phase. Experimentally, it can be created through a protocol described in detail in Refs.~\cite{wimmer2013optical, martinthesis}. Figure~\ref{expan1}(b) then shows the numerical differential intensity resulting from the application of a realistic narrow defect, with $\sigma_n\!=\!2$, $\sigma_m\!=\!2$, $\varphi_0=0.01$, $m_d=11$ and $n_d=0$. As can be seen, this propagation pattern appears to be similar to that on top of a uniform light field [see e.g. Fig.~\ref{cones}], with excitations at large positions suppressed by the drop-off in the intensity of the cloud. 
	
	Despite this apparent similarity, the expanding light field does have two very important consequences for the propagation of sound waves. Firstly, the light field expansion will drag perturbations with it, artificially inflating the measured excitation speed. Secondly, in the absence of gain or loss, the local intensity of the light field will drop over time, causing the effective local speed of sound to vary. We now show that both of these effects can be accounted for by applying a numerical procedure to estimate the local expansion speed of the light field. We previously applied this procedure to analyse the experimental data in~\cite{experiment}, as well as associated numerical simulations. As mentioned above, in that case, the experimental data could not be quantitatively compared to the analytical speed of sound due to experimental uncertainties, for example, in determining the value of the experimental nonlinearity parameter. In what follows, we provide a general theoretical framework for this physics.
	
	As above, we first fit cuts of the differential intensity at each time step using Eq.~\ref{eq:fit}; this is shown, for example, for $m=100$ in Fig.~\ref{expan1}(c), where the blue points are the numerical data and the black curve is the fit. We extract the positions of the innermost intensity minima by finding the minima of the fits, with the results plotted in Fig.~\ref{expan1}(d) as a function of time. As we want to allow for the possibility that the speed of the intensity minima varies over time, we perform a moving linear fit to the data, which gives us an estimate for the instantaneous speed of the minimum, $v_M$ shown by red data points in Fig.~\ref{expan1}(e). Here, we have chosen the moving fit for a time step $m$ to be taken over eleven time steps centred around $m$. As can be seen, the estimated speed is (i) not a constant as a function of time step $m$, and (ii) significantly larger than the expected analytical speed of sound for the initial nonlinearity~\eqref{eq:sound}, plotted in (e) with a blue dashed line. To explain these observations, we need to take into account the expansion of the light field over time.

	\begin{figure*}[ht]
		\includegraphics[width=1\linewidth]{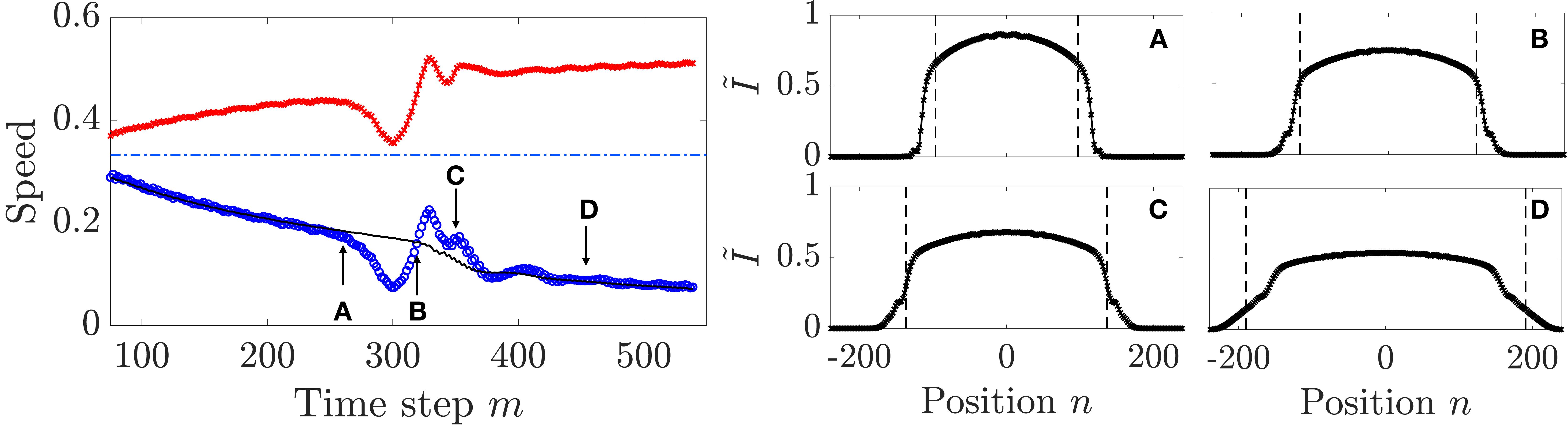}
		\caption{({\it Left panel}) Results from Fig.~\ref{expan1}(f) now plotted for longer times, with red data points indicating $v_M$; blue data points indicating the numerically extracted $v_s = v_M -v_E$; blue dotted-dashed line indicating the speed of sound $v_s^0$ corresponding to the initial nonlinearity; the black solid line indicating the analytical prediction for the local speed of sound $v_s ( n_{\text{min}})$. As can be seen, there is excellent agreement between the numerical estimate and the analytical prediction (blue vs. black), except for the $250\lesssim m \lesssim 400$ interval. ({\it Right panels}) To understand this, we plot the intensity of the unperturbed optical field $\tilde{I} = (|u_n|^2 + |v_n|^2)_\text{unpert.}$ at different times as indicated in the left panel. The vertical dashed lines indicate the positions of the minima at each time. As can be seen, the deviations visible in the left panel appear where the wave emanating from the defect hits the shoulder of the intensity profile where the local intensity decreases sharply. 
		}\label{expan2}	
	\end{figure*}

	As discussed in Ref.~\cite{experiment}, our numerical practice
	protocol for estimating the expansion speed of the light field is based on a version of the 1D continuity equation:
	\begin{eqnarray}
		\frac{d I_A}{  d t} = -v_E\frac{I (A)}{2}  + \beta I_A \label{eq:cont}
	\end{eqnarray}
	where $v_E$ is the drag speed, $\beta$ is an overall linear amplification factor and $I_A = \sum_{n \in \mathcal{A}} I(n)$ 
	is the integrated local intensity, $I (n) = |u_n|^2 + |v_n|^2$, in the absence of the defect. The summation domain $\mathcal{A}$ is chosen as $-\infty \!\leq\! n\!\leq\!  A$ when $A>0$ or $A\! \leq\! n\!\leq\!  \infty$ when $A<0$; this ensures that the integral always includes at least half of the light field to minimize the effects of noise. Physically, Eq.~\eqref{eq:cont} balances the rate of change of the intensity within a selected region to the amount of light flowing in/out and the net gain within that region. The factor of $1/2$ in the first term on the right-hand side of the equation takes into account that the lattice spacing is $2$ in this system, and so that the local density of light is given by ${I (A)}/{2}$. In the optical mesh lattice, time is discretized into discrete time steps and we can approximate this continuity equation as:
	\begin{eqnarray}
		\frac{ I_A (m) - I_A (m- \Delta m) }{  \Delta m }=-v_E\frac{I (A)}{2} + \beta I_A (m) \label{eq:num}
	\end{eqnarray}
	where we have included explicitly the dependence on the time step. To determine the linear amplification factor $\beta$, we can apply this equation to the summed intensity over the entire system:
	\begin{eqnarray}
		\frac{ I_\infty (m) - I_\infty (m- \Delta m) }{  \Delta m }= \beta I_\infty (m).
	\end{eqnarray}
	Numerically, we choose to set $\beta=0$, but experimentally, $\beta$ can be non-zero due to intrinsic loss and gain in the system~\cite{experiment}. 
	Putting this back into~\eqref{eq:num}, the numerically-extracted instantaneous drag speed, $v_E$, at the position of the minimum, $n_{\text{min}}$, can be calculated. The result for $\Delta m =4$  is shown  as black dots in Fig.~\ref{expan1}(e). Note that $n_{\text{min}}$ is obtained as the minimum of a fitting function and so can take continuous values, while the drag speed can only be calculated at the available lattice sites of the optical mesh lattice, making this an approximation of the speed. In any case, as expected, the drag speed increases over time, corresponding to the intensity minima moving outwards into faster-moving regions of the light field.

	\begin{figure}[ht]
		\includegraphics[width=1\linewidth]{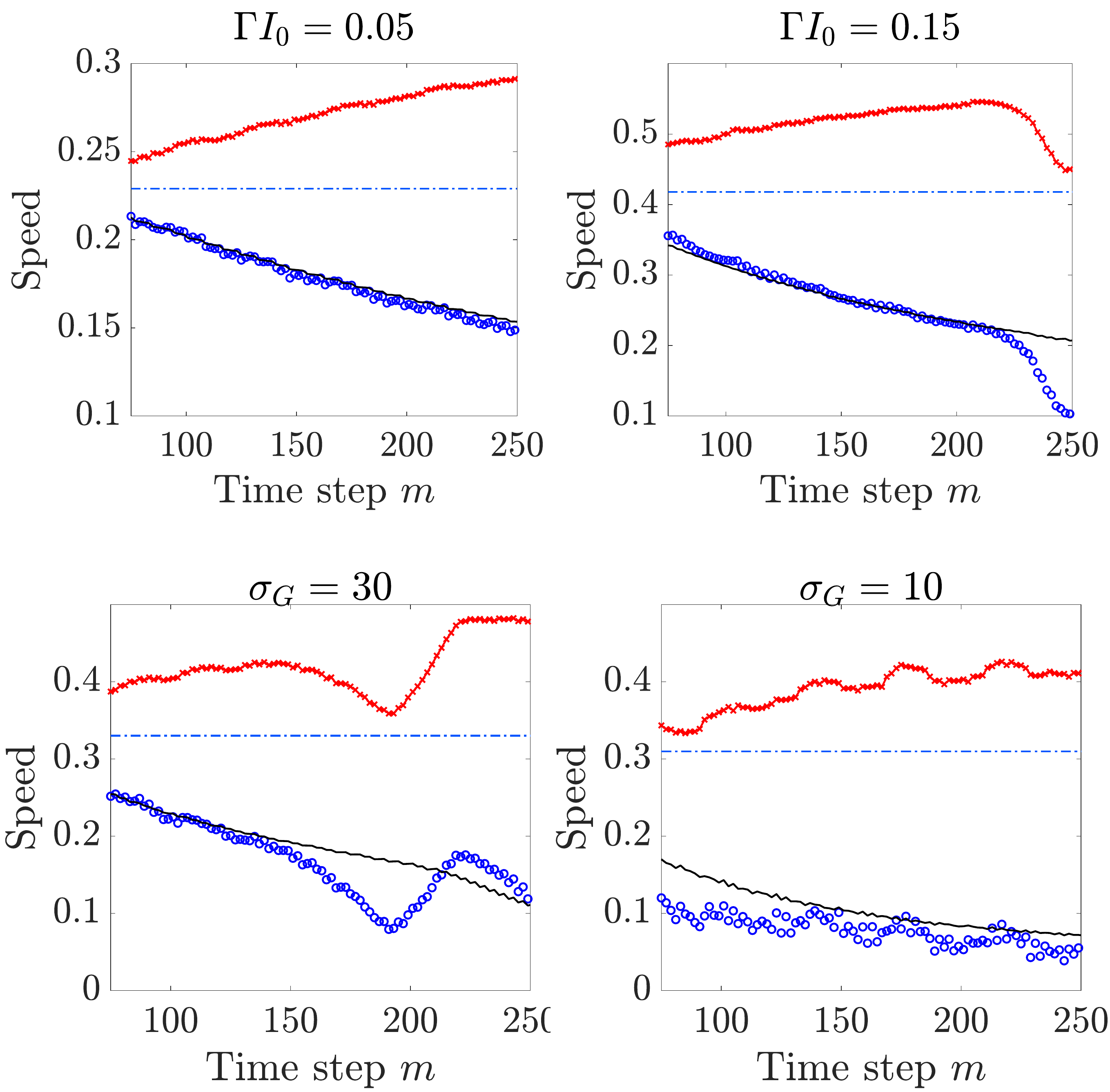}
		\caption{({\it Top Row}) Plot of the different speeds as in Fig.~\ref{expan2}~(a) but for two different values of the nonlinearity parameter $\Gamma I_0=0.05$ and $0.15$ (instead of $0.1$). Again there is very good agreement between the numerically-estimated speed of sound (blue dots) and the analytical prediction for the local speed of sound $v_s ( n_{\text{min}})$ (black line). The deviation between these two again occurs over an interval where the defect hits the shoulder of the spatial intensity profile  (as in Fig.~\ref{expan2} (b)); as the expansion rate of the optical field grows with the nonlinearity, the time at which the discrepancy occurs correspondingly decreases. ({\it Bottom Row}) The same plot for $\Gamma I_0=0.1$ but narrower initial Gaussian profile widths $\sigma_G=30$ and $10$ (instead of $50$). As the initial width of the optical field decreases, the defect hits the shoulder of the intensity profile at an earlier time, leading to an earlier deviation between the estimated and local speed of sound. For very narrow clouds, noise in the numerical estimates is more apparent as the defect quickly leaves the central region of high intensity and the very concept of local speed of sound becomes inaccurate. Nevertheless, a good qualitative agreement can still be observed. 
		}\label{expan3}	\end{figure}

	Importantly, we see from Fig.~\ref{expan1}(e) that the drag speed $v_E$ accounts for a significant fraction of the observed speed of the intensity minimum, $v_M$. In Fig.~\ref{expan1}(f), this is shown explicitly by plotting the estimated speed of sound $v_s = v_M - v_E$, which turns out much lower than $v_M$ and which decreases over time. This estimation for $v_s$ is also much lower than the analytically-expected result~\eqref{eq:sound}  for the initial nonlinearity, $\Gamma I_0=0.1$ [blue dashed line in (e)]. This is easily explained in terms of the drop of the local intensity at the position of the defect over time, which leads to a decreasing effective speed of sound. For a quantitative analysis, we can extract the average local intensity in each loop at the position of the minimum $I( n_{\text{min}})/2$, and use this value  within a sort of local density approximation to calculate a local instantaneous speed of sound, $v_s ( n_{\text{min}})$ via~\eqref{eq:sound}. The result is plotted with a solid black line in Fig.~\ref{expan1}(f) and is indeed in excellent agreement with the numerical data for all times of experimental interest. This provides a solid numerical confirmation of the protocol used to extract the speed of sound in the recent experiment~\cite{experiment}.
	
	Differently from experiments, we can extend our numerical analysis out to even later times as shown in Fig.~\ref{expan2}. There we see that marked deviations do occur between the numerical and the analytical local instantaneous speed of sound around $250\lesssim m \lesssim 400$. To understand this feature, in the right panels (A-D) we plot the corresponding spatial intensity profiles of the optical field in the absence of the defect. Here, vertical dashed lines indicate the positions $\pm n_{\text{min}}$ of the innermost minima in the intensity profiles in the presence of the defect. From these plots, we infer that the deviations occur when the sound waves move through the shoulder of the optical field, where the intensity drops sharply as a function of position and the local density approximation that implicitly underlies $v_s(n_{\textrm{min}})$ is no longer justified. On the other hand, agreement between these estimates is excellent in all other cases where the local intensity of the underlying optical field is sufficiently slowly varying, even after the defect has passed through the shoulder of the spatial intensity profile. 
	
	As further checks, we repeat the calculation for different values of the initial nonlinearity parameter, $\Gamma I_0$, as shown in the top panels of Fig.~\ref{expan3}. The numerical estimate of the local speed of sound is again very accurate except when the perturbation is crossing the sharp shoulder of the spatial intensity profile. This shoulder is reached earlier for higher nonlinearities due to the higher sound velocity, which leads to an earlier onset of the deviations. We can also vary the initial Gaussian width of the optical field, $\sigma_G$, as shown in the bottom panels of Fig.~\ref{expan3}. Decreasing the size of the initial field leads again to an earlier onset of the deviation window, as less time is needed for the defect to reach the shoulder of the spatial intensity profile. 
	
	Note that in the recent experiment~\cite{experiment}, the Gaussian width and maximum time step were respectively $\sigma_G=8$ and $m_{\textrm{max}}=110$, due to the number of round-trips required to generate the initial Gaussian field before the start of the superfluidity experiment~\cite{wimmer2013optical, martinthesis}. Further numerical calculations with these experimentally-realistic parameters were analysed in the Supplemental Material of Ref.~\cite{experiment}. In the future, it would be interesting to realize more roundtrips, which would leave more time for the creation of the initial field and thus allow for the creation of wider Gaussian light fields and hence more accurate experimental measurements of the local speed of sound.
	
	

\section{Superfluidity properties
} \label{sec:landau}

The characterization of collective excitations and of their propagation speed reported in the previous Section is the natural  starting point to investigate the superfluidity properties of the fluid of light in a nonlinear optical mesh lattice. In this Section, we summarize our advances in this direction, based on an extension of the Landau criterion for superfluidity~\cite{pitaevskii2016bose, CastinLectures,RevModPhys.85.299} to our specific spatio-temporally periodic geometry.

The Landau criterion, in its simplest formulation, provides a general and intuitive way to predict the breakdown of superfluid behaviour when a fluid system is traversed by a uniformly moving impurity at sufficiently large speed. This simple criterion amounts to deriving a critical velocity: 
\begin{eqnarray}
	v_c = \text{min}_k \left(\frac{ \theta (k)}{k}\right), \label{eq:criticalvelocity}
\end{eqnarray}
below which a weak moving perturbation is not able to generate collective excitations in the fluid, which then behaves as a frictionless superfluid.
Above this critical velocity collective excitations start being generated in the fluid and dissipations sets in. 

For the simple case of a three-dimensional, homogeneous BEC with short-range interactions described by the GPE (see Eq.~\ref{eq:gpedispersion}), it can be shown that the critical velocity is directly given by the speed of sound: $ v_c \!=\! v_{\text{GPE}}\! =\! \sqrt{g I_0 / M}$~\cite{pitaevskii2016bose, CastinLectures}. This means that the breakdown of superfluidity is associated with the emission of long-wavelength ($k\rightarrow0$) sound waves, and that superfluidity vanishes altogether in the limit of a non-interacting Bose gas, where $v_c=v_{\text{GPE}}= 0$. In other systems, the emergence of additional features in the Bogoliubov dispersion significantly affects the superfluid response; for example, the Bogoliubov dispersions of superfluid helium or a dipolar BEC can both exhibit a so-called ``roton" minimum at a finite momentum which then sets the critical velocity to a lower value than the speed of sound~\cite{pitaevskii2016bose}. In polariton light fields, there can also be important non-equilibrium effects, leading to the emergence of novel scattering regimes~\cite{Wouters:PRL2010}. 

In this section, we explore the effect of a moving defect in an optical mesh lattice, firstly by considering suitable generalizations of the Landau criterion in Sec.~\ref{sec:analyticslandau}, and secondly from numerical simulations in Sec.~\ref{sec:numericslandau}. As a main result of our study, we find that for any lattice, a strict application of the Landau criterion would predict a vanishing critical velocity, so that a moving defect would always excite the light field~\cite{experiment}. However, in practice, this prediction is strongly modified when the spatial defect width is increased, so as to reduce scattering at higher momentum. In this regime, we observe in fact that the response of the optical field is effectively ``superfluid-like": a characteristic threshold for the onset of dissipation is apparent, which depends on the nonlinearity as expected from a restricted Landau criterion, and which observably drops below the speed of sound for high enough nonlinearities.

\subsection{Umklapp processes} \label{sec:analyticslandau}

The Landau criterion applies the principles of energy and momentum conservation to the Bogoliubov excitation spectrum of a system~\cite{pitaevskii2016bose, CastinLectures}. For example, if we have a defect moving at a constant speed, $s$, (as realised e.g. by taking an $m$-dependent $n_d = s (m - m_d) $ in Eq.~\ref{eq:defectform}), then a simple way to visualise the Landau criterion is by drawing  on top of the excitation spectrum a straight line passing through the origin with a slope $s$. In continuous space and time, the excitation processes that are allowed by energy and momentum conservation correspond to the intersections of this line with the Bogoliubov dispersion~\cite{Carusotto2013}. In particular, the critical velocity, $v_c$, is then read off as the speed at which the straight line first touches the Bogoliubov bands from below. If the system is a superfluid, then $v_c \neq 0$, and for $s<v_c$, there are no resonant excitations and, thus, no dissipation.

\begin{figure*}[ht]
	\includegraphics[width=1\linewidth]{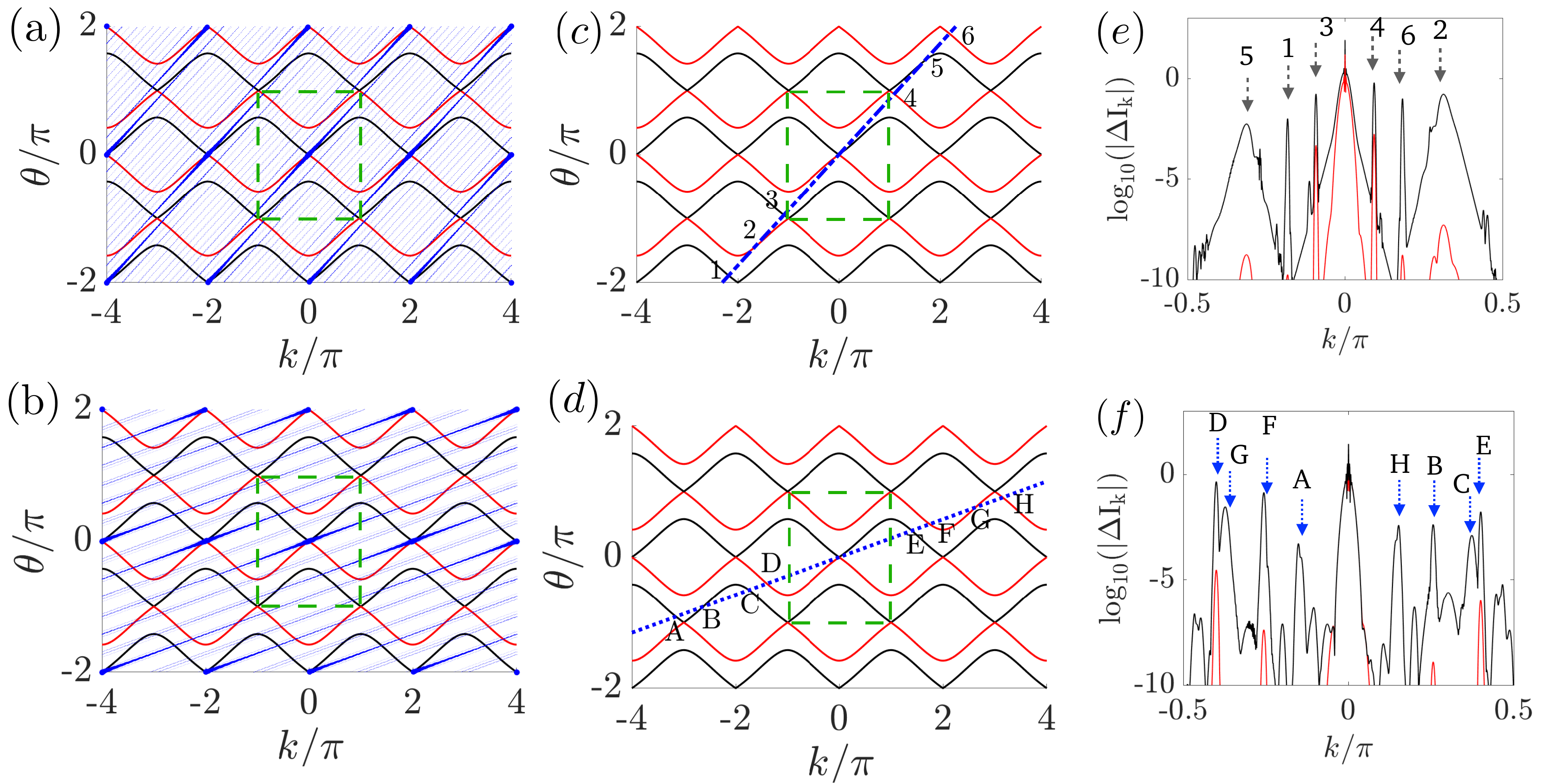}
	\caption{ (a) \& (b) The Bogoliubov dispersion for $\Gamma I_0 =0.25$ in an extended Floquet-Bloch zone scheme, with the range shown in Fig.~\ref{bogdispersion} outlined by a green dashed box. Also plotted are the straight lines (\textit{blue  points}) corresponding to a defect moving at  (a) $s=1.5 v_{\rm{sound}}$ and (b) $s=0.5 v_{\rm{sound}}$. Due to the spatial and temporal periodicities of the optical mesh lattice, there are an infinite number of replicas of the defect straight lines. However, as discussed in the main text, for a sufficiently slowly varying defect, the corresponding Fourier-space defect potential is highly peaked around certain momenta for each line and becomes small elsewhere; this relative importance is schematically represented by an increased size of the blue points, with the largest weight being concentrated around $k=0$ and equivalent points in the extended Floquet-Bloch scheme. (c) \& (d) The dispersion as in (a) \& (b) respectively, but only with the fundamental $r=r'=0$ defect line (now unadjusted for the relative defect weight), in order to highlight the first intersections between these lines and the Bogoliubov bands as labelled by numbers and letters respectively. (e) \& (f) The logarithm of the absolute value of the differential intensity spectra (Eq.~\ref{eq:iq}), as calculated at the late time step $m=900$ for a defect with spatial width $\sigma_n=0.25$ ({\it black line}) and $\sigma_n=0.75$ ({\it red line}) moving through a uniform light field with  $\Gamma I_0 =0.25$ at the speed values (e) $s=1.5 v_{\rm{sound}}$  and (f) $s=0.5 v_{\rm{sound}}$. The defect parameters are:  $\sigma_m=100$, $m_d=500$ and $\varphi_0=0.01$. Labelled arrows indicate the momenta of the lowest intersections shown in (c) \& (d) back-folded into the range: $-\pi/2 \leq k < \pi/2$. }\label{floquet}	
\end{figure*}

The situation is slightly more complicated in our spatially and temporally periodic optical mesh lattice configuration, as illustrated schematically in Fig.~\ref{floquet}(a) and (b). Given the discrete spatial and temporal periodicities, one needs in fact to work in the extended-zone scheme and include replicas of the straight line according to these periodicities. As a result, a direct application of the Landau criterion leads to a striking conclusion: the critical velocity is always zero. This can be seen immediately from the above geometric argument by realising that the aforementioned straight lines always intersects with the Bogoliubov dispersion of the optical mesh lattice in some Brillouin zone, see e.g. Fig.~\ref{floquet}(a) and (b). On this basis, one may expect that a lattice system can never respond like a true superfluid to a moving defect, as the defect always leads to dissipation. This notwithstanding, we note that such systems can still display other typical features of superfluidity, for example for what concerns the robustness of a superflow to dynamical and energetic instabilities~\cite{wu1,wu2,Menotti_2003,modugno} or the rigidity of the system under ``twisted boundary conditions”~\cite{Rey}. 

Furthermore, the above argument based on the Landau criterion does not take
into account the momentum-selectivity of the excitation process due to the width of the defect, $\sigma_n$. In practice, excitations at higher momenta $k\gg 1/\sigma_n$ are suppressed in the case of a wide defect. To see this, let us consider a general lattice defect potential moving at a constant speed with the form $V(n-sm, m) =  \beta(m) \alpha(n-sm)$, where $\alpha(n-sm)$ corresponds to the moving spatial profile of the defect and $\beta(m)$ to the temporal profile. For example, a stationary defect of the form Eq.~\ref{eq:defectform} with a $m$-independent position $n_d$ is captured by this functional form with Gaussian-shaped $\alpha(n)$ and $\beta(m)$ and $s=0$. Imposing periodic boundary conditions, the discrete Fourier transform of the moving defect potential can be defined as:
\begin{eqnarray}
	\tilde{V} (k , \theta) &\equiv& \sum_{m=0}^{M-1} \sum_{n=0}^{N-1} V (n-sm,m) e^{ i \frac{2\pi}{M} \varepsilon m} e^{- i \frac{2\pi}{N} q n} \nonumber \\
	&=& \sum_{m=0}^{M-1} \beta(m) e^{ i \frac{2\pi}{M} \varepsilon m} \sum_{n=0}^{N-1}  \alpha(n-sm) e^{- i \frac{2\pi}{N} q n} \quad
	\label{eq:intermediate_V}
\end{eqnarray}
where $N$ denotes the number of sites along $n$, and $M$ that along $m$, with $\varepsilon$ and $q$ being integers that run from zero up to $M-1$ and $N-1$ respectively and which are related to a (discretized) energy and momentum as $\theta  = ( 2 \pi / M ) \varepsilon $ and $k = (2 \pi / N ) q $ respectively. Continuum variables are recovered in the limit that $M$ and $N$ tend to $\infty$. 

To isolate the dependence on $m$ in the second part of the equation, we can apply Poisson's summation formula 
\begin{equation}
	\sum_{n=-\infty}^{\infty}\,f(n)=\sum_{r=-\infty}^{\infty}\,\bar{f}(2\pi r)\,,
\end{equation}
where $\bar{f}(k)=\int_{-\infty}^\infty\,dx\,f(x)\,e^{-ikx}$ is the standard {\em continuous} Fourier transform of $f(x)$, to the second part of the equation. This gives
\begin{multline}
	\sum_{n=-\infty}^{\infty}\,\alpha(n-sm)\,e^{-i\frac{2\pi}{N}qn}\\ =\sum_{r=-\infty}^{\infty}\,\bar{\alpha}\left(\frac{2\pi}{N}q+2\pi r\right)\,e^{-i(\frac{2\pi}{N}q+2\pi r)sm}
\end{multline}
where $\bar{\alpha}(k)$ is the continuous Fourier transform of $\alpha(j)$.

Substituting this formula back into the full expression \eqref{eq:intermediate_V} where the sums have been extended to infinity to account for the $N,M\to \infty$ limit, we obtain:
\begin{multline}
	\tilde{V} (k , \theta) = \sum_{m,r=-\infty}^{\infty} \beta(m) e^{ i \frac{2\pi}{N} m (\frac{N}{M}\varepsilon - (q+N r) s )} \,\tilde{\alpha}(k+2\pi r) \\
	= \sum_{r=-\infty}^{\infty}\,\tilde{\beta} \left(\theta - 2\pi r s -k s \right) \bar{\alpha}(k + 2\pi r)  \\
	= \sum_{r,r'=-\infty}^{\infty}\,\bar{\beta} \left(\theta - 2\pi r s -k s -2\pi r' \right) \bar{\alpha}(k + 2\pi r) 
	\label{eq:defectfourier}
\end{multline}
where $\tilde{\beta}(\theta)$ and $\bar{\beta}(\theta)=\int_{-\infty}^\infty\,dt\,\beta(t)\,e^{i\theta t}$ are the  discrete and continuous Fourier transforms of $\beta(m)$, respectively.

The $r'\neq 0$ replicas separated by $\Delta \theta=2\pi$ stem from the effective Floquet periodicity of our configuration.
The $r\neq 0$ replicas separated by $\Delta k=2\pi$ correspond to the Bragg momentum of the lattice: in the time domain, the corresponding frequency side-bands can be physically understood as the motion of the defect making the effective potential periodically either overlap with some site or fall in the void between sites, with temporal frequency $\Delta \theta=2\pi s$.

For slowly varying potentials $\sigma_m\gg 1$, the Fourier-space defect potential $\tilde{V}(k,\theta)$ is concentrated along an infinite series of parallel straight lines $\theta=(k+2\pi r)s+2\pi r'$ corresponding to the usual defect dispersion and its replicas, as displayed in Fig.~\ref{floquet}(a) and (b)~\footnote{Note that this reasoning was carried out assuming a standard square spatio-temporal lattice in the Fourier transform in (\ref{eq:intermediate_V}). Inclusion of all subtleties of the optical mesh lattice band dispersion as discussed in Sec.\ref{sec:bog} would introduce additional straight lines passing through, e.g., $k/\pi=1$ and $\theta/\pi=1$ restoring the full symmetry of the diagram. A complete discussion of these issues goes however beyond our discussion and is of no relevance for our conclusions here.}. For a given $r,r'$ line, the defect weight will also be controlled by $\bar{\alpha}  ( k + 2 \pi r)$, which reaches a maximum when $k=-2\pi r$. This maximal weight is indicated schematically in Fig.~\ref{floquet}(a) and (b) by increasing the size of the markers (which compose the straight lines) around $k\approx -2\pi r$. As can be seen, the defect weight is most significant around the center of the green dashed box, and, due to the replicas, around equivalent points in the extended Floquet-Bloch zone.

As the argument, $ |k + 2 \pi r|$, quickly becomes large for Umklapp excitations, such components will only be visible for sufficiently narrow defects such that the Fourier transform of the defect potential has a significant weight at these large argument values. For wide-enough defects with $2 \pi \sigma_n \geq 1$, the Landau criterion restricted to a small momentum/frequency range at the center of the Brillouin zone  $-\pi < k,\theta  < \pi$ 
can instead be expected to provide a good description of the system behaviour. 
We can verify the validity of this argument numerically, by simulating the constant motion of spatially-narrow defects (taking $n_d = s (m - m_d) $ in Eq.~\ref{eq:defectform}) across a uniform light field. To minimise additional effects arising from turning on and off the defect, we choose a large temporal defect width $\sigma_m = 100$ and a small defect amplitude $\varphi_0=0.01$. In order to identify the excitation of higher Bogoliubov bands, we first replot in Fig.~\ref{floquet}(c) and (d) the dispersion from (a) and (b), but now showing only the corresponding fundamental  $r=r'=0$ line so that we can clearly indicate, with numbers and letters respectively, the first expected intersections between the defect dispersion and the Bogoliubov bands. We then plot in Fig.~\ref{floquet} (e) and (f) the logarithm of the absolute value of the differential intensity spectrum~\eqref{eq:iq} as a function of the momentum with respect to the effective 1D lattice; this is calculated numerically in each case for a spatial width of $\sigma_n=0.25$ ({\it black line}) and $\sigma_n=0.75$ ({\it red line})   at the late time step $m=900$ with $m_d=500$. As can be seen, the largest peak occurs at $k=0$, signifying the depletion of the light field into Bogoliubov quasi-particles and  the residual excitation of low momenta sound waves. The momenta of the next largest peaks in the spectra are all in excellent agreement with the positions of the lowest intersection points highlighted in Fig.~\ref{floquet}(c) and (d), and indicated in Fig.~\ref{floquet} (e) and (f) by labelled arrows. These higher Floquet-Umklapp excitations are much weaker than the low-energy excitations at $k=0$, and they get suppressed when the spatial width of the defect $\sigma_n$ is increased.

\begin{figure}[!]
	\includegraphics[width=0.9\linewidth]{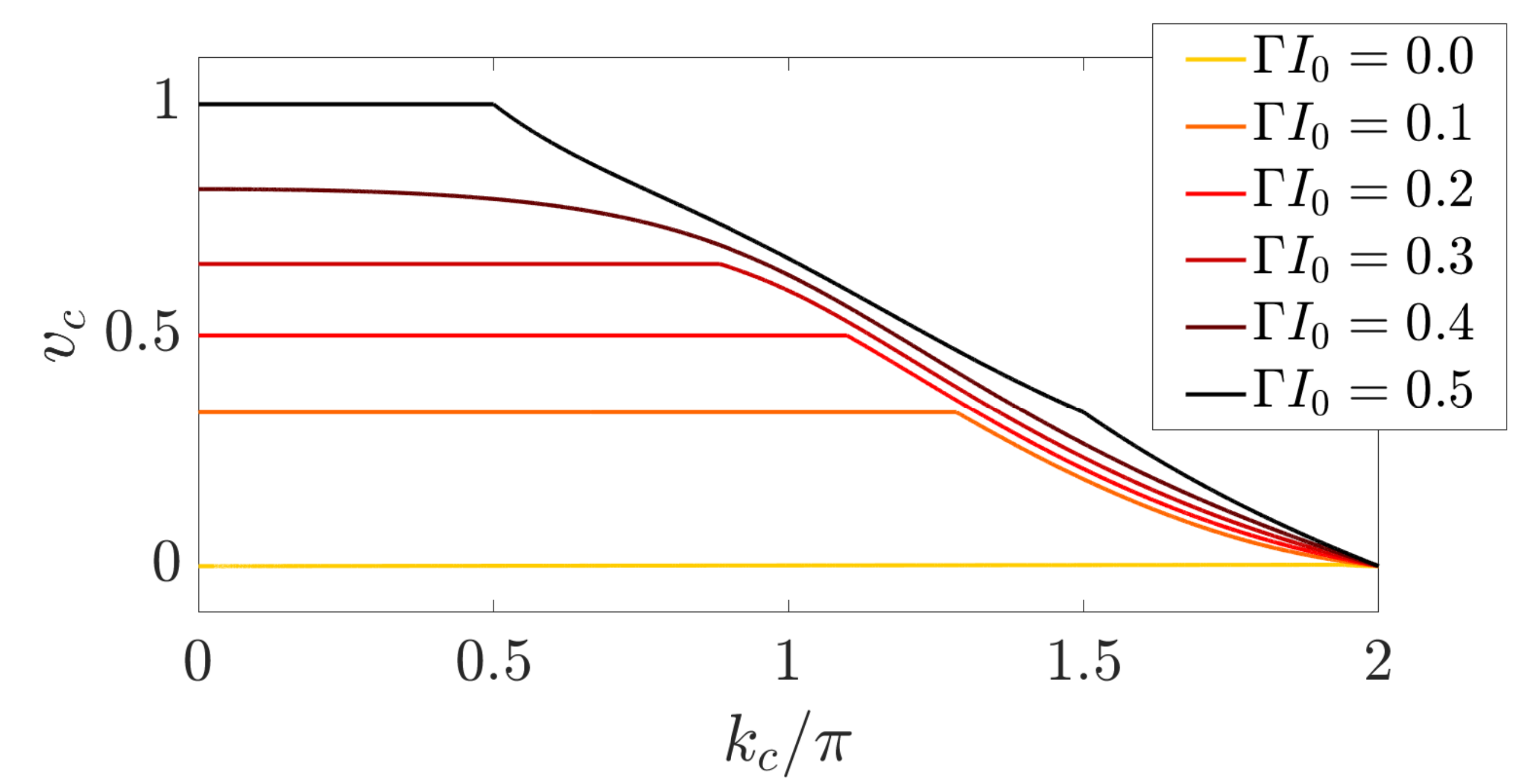}
	\caption{ The critical velocity as calculated from the Bogoliubov dispersion within a region $-k_c \leq k \leq k_c$. For $k_c= 2 \pi$ or larger, $v_c =0$ for all stable nonlinearities, indicating that the system is not superfluid. However, for lower values of $k_c$, a non-vanishing $v_c \neq 0$ is found in the presence of a finite nonlinearity, indicating that  ``superfluid-like" behaviour may be observed for sufficiently wide defects. }\label{landaucrit}
\end{figure}

As Umklapp excitations are therefore effectively strongly-suppressed for wide-enough defects, we may expect that restricting the Landau criterion to a limited momentum range may well-describe the system behaviour. In Fig.~\ref{landaucrit}, we plot the critical velocity that would be predicted by applying the Landau criterion to the Bogoliubov dispersion within a restricted region $-k_c \leq k \leq k_c$. 
In the linear regime, the critical velocity is always vanishing independently of the cut-off $k_c$, as superfluidity is a nonlinear phenomenon. For $k_c= 2 \pi$ or larger (i.e. for a restricted region equal to twice the Brillouin zone shown in Fig.~\ref{bogdispersion} or larger), the critical velocity also vanishes for all stable nonlinearities, indicating that the system is never a true superfluid. However, for lower values of $k_c$, the critical velocity is non-zero in the presence of a nonlinearity, with $v_c \rightarrow v_{\text{s}}$ as $k_c \rightarrow 0$, as it happens in simple BECs. This means that the system may appear to be ``superfluid-like" with a relative suppression of dissipation. 

To confirm that such behaviour can be indeed observed in practice, we now numerically simulate the motion of moving defects in a nonlinear optical mesh lattice first for uniform and then for expanding light fields, comparing our results with the restricted Landau criterion. 

\subsection{Superfluid-like response of uniform light fields} \label{sec:numericslandau}

We first investigate the superfluid-like response of a stationary and uniform optical field. To this purpose, we again apply a Gaussian phase defect, as discussed in Sec.~\ref{sec:measure}, but we now move the defect by changing the peak defect position, $n_d$ as a function of the time step $m$ at a constant speed, $s$, across the uniform optical field, according to $n_d = s (m - m_d) $. We shall then measure how much the defect excites the system as a function of the defect speed.

\begin{figure}[!]
	\includegraphics[width=0.95\linewidth]{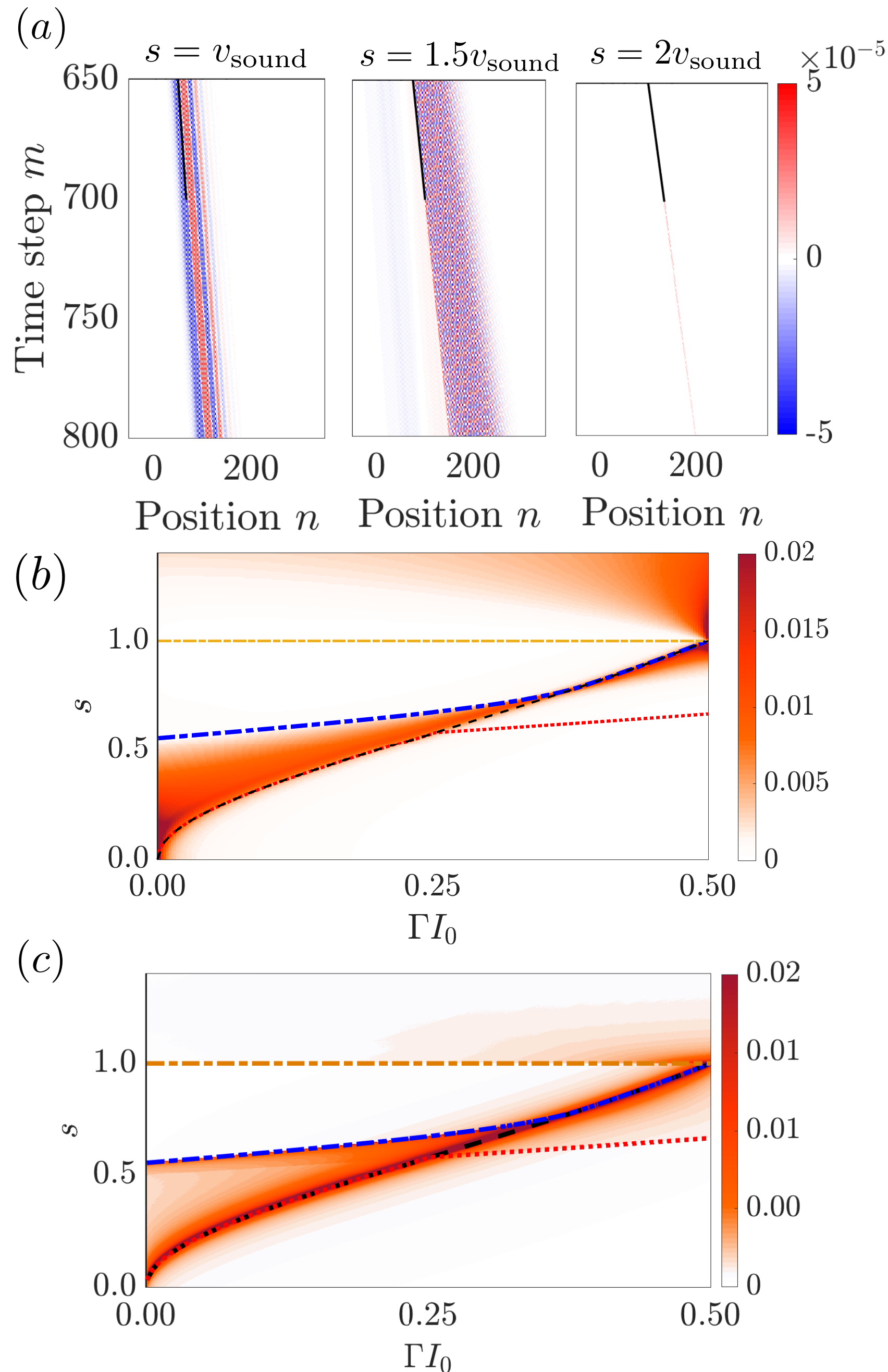}
	\caption{(a) Spatio-temporal colorplot of the numerical differential intensity (Eq.~\ref{eq:diff}) for a temporally smoothly-varying, spatially narrow and very weak defect ($\sigma_m=200$, $\sigma_n=1$, $\varphi_0=0.00001$, $m_d=500$, $n_d=0$), moving at speeds $s=\{1, 1.5, 2\} v_{\rm{sound}}$ as indicated by the black parallelogram across a uniform light field, with nonlinearity $\Gamma I_0 =0.1$. (b) The total emission intensity (Eq.~\ref{eq:observable}) plotted as function of the defect velocity and nonlinearity for  $m_{\rm{max}}=1000$ and $j=101$.  As a guide to the eye, we have marked the speed of sound ({\it black dashed line}); lower and upper critical velocities for the first Bogoliubov band ({\it red dotted line} and {\it blue dashed line} respectively); the lower critical velocity for the second Bogoliubov band ({\it yellow dashed line})  within $-\pi \leq k,\theta < \pi$. (c) The fraction of states (Eq.~\ref{eq:fraction}) in the Bogoliubov bands which approximately satisfy the restricted Landau criterion as calculated with $N_T=15000$ and $\delta \theta=0.001$. Other marked lines are the same as in panel (b).}\label{narrow}	
\end{figure}

\subsubsection{Narrow defect}

In Fig.~\ref{narrow}, we start with the case  of a 
spatially very narrow defect. Such a defect is able to excite
higher-momentum perturbations. The numerically computed spatio-temporal differential intensity is plotted in Fig.\ref{narrow}(a)  for three different values of the defect speed. As can be seen, the excitation pattern is very complicated.
To quantify the total intensity of the emission from the defect, it is then useful to introduce the observable: 
\begin{eqnarray}
	\tilde{I}
	= \frac{1}{j} \sum_{m =m_{\rm max}-j+1}^{m_{max}} \ \left[ \sum_n |\Delta I (n,m) |  \right]\label{eq:observable}
\end{eqnarray}
which temporally averages the spatially integrated  differential intensity over the last $j$ time steps of the propagation. 

If we plot the total emission intensity~\eqref{eq:observable} as a function of the speed $s$ and the nonlinearity $\Gamma I_0$, 
then a clear structure is observed as significant emission only occurs within certain ranges of defect speeds [Fig.\ref{narrow}(b)], as would be expected for a standard superfluid. 
Outside this range of $s$, the defect cannot efficiently excite any perturbations in the first (i.e. innermost) Bogoliubov band, as all resonant excitations happen to be at too large momenta even for the narrow defect considered here: 
only excitations within $-\pi \leq k < \pi$ are significant and all emission lies within the window predicted by applying the Landau criterion over this limited momentum range. 

The thickness of the orange stripe in Fig.\ref{narrow}(b), indicating a sizable total emission, depends on the curvature of the lowest Bogoliubov band, which is positive for small $\Gamma I_0$, then negative at large $\Gamma I_0$. As a guide to the eye, we have indicated the speed of sound (black dashed line); the lower and upper critical velocities for the first Bogoliubov band (red dotted line and blue dashed line respectively). As can be seen, at low nonlinearities, the speed of sound and the lower critical velocity of the first band coincide, and the emission is concentrated between this and the upper critical velocity.
The relatively flat distribution of the emission intensity within the stripe is related to the velocity-independence of the friction force in one-dimensional superfluids above the critical speed~\cite{Astrakharchik:PRA04}.
In contrast, at higher nonlinearities, the speed of sound coincides with the upper critical velocity of the first band, and so emission from the first band is concentrated below this speed. For the sake of completeness, the lower critical velocity for the second Bogoliubov band has been shown in Fig.\ref{narrow}(b) as a yellow dashed line: except for a small region at high nonlinearities $\Gamma I_0\lesssim 0.5$, emission into this band is suppressed by the small value of the excitation matrix element.

Further insight on this physics can be obtained by comparing the total emission intensity shown in Fig.~\ref{narrow}(b) to a numerical estimate for how many Bogoliubov states approximately satisfy our restricted Landau criterion as a function of the optical nonlinearity and defect speed. As discussed above, a simple way to visualise the restricted Landau criterion is to ask where the defect dispersion $\theta=ks$ intersects with the Bogoliubov bands with $k< k_c =\pi$. To numerically estimate what fraction of available states may satisfy this criterion, we discretize the momentum, $k$, into $N_T$ equally-spaced values over the range $k\in [0, \pi]$, and then calculate the associated set of discrete Bogoliubov energies ${\Theta} (k)$ for the two bands assuming $0\leq \theta < \pi$ in Eq.~\ref{eq:bog}. We then define the fraction of these states which approximately satisfy the restricted Landau criterion as:
\begin{eqnarray}
	\xi = \sum_{k<k_c} \frac{N[|\Theta (k) - k s | < \delta \theta] }{2 N_T} \label{eq:fraction}
\end{eqnarray}
where the factor of 2 in the denominator occurs because here we consider the two lowest-energy Bogoliubov bands and where $N[|\Theta (k) - k s | < \delta \theta] $ denotes the number of discrete Bogoliubov states for which the energy lies within a small energy window $\pm \delta \theta$ of the defect dispersion. The small parameter $\delta \theta$ helps compensate for the discretization, but is also physically motivated by the finite energy-width expected for a defect switched on for a finite time [c.f. Eq.~\ref{eq:defectfourier}]. 

In Fig.~\ref{narrow}(c), we plot Eq.~\ref{eq:fraction} for $N_T=15000$ and $\delta \theta=0.001$ (which is on the order of the energy-width expected for a Gaussian defect with $\sigma_m=200$), with the same marked lines as shown in Fig.~\ref{narrow}(b). As can be seen, the largest fraction of available Bogoliubov states is around the speed of sound (black dashed line), where strong emission is also observed numerically in Fig.~\ref{narrow}(b). Note that the fraction of states in Eq.~\ref{eq:fraction} also predicts other features like a significant emission around the upper critical velocity for the first Bogoliubov band (blue dashed line), which are not observed numerically in Fig.~\ref{narrow}(b); this is because Eq.~\ref{eq:fraction} does not take into account other effects, such as the finite spatial width of the defect which suppresses larger-momentum excitations nor the $k$-dependence of the matrix elements.

\begin{figure}[!]
	\includegraphics[width=1\linewidth]{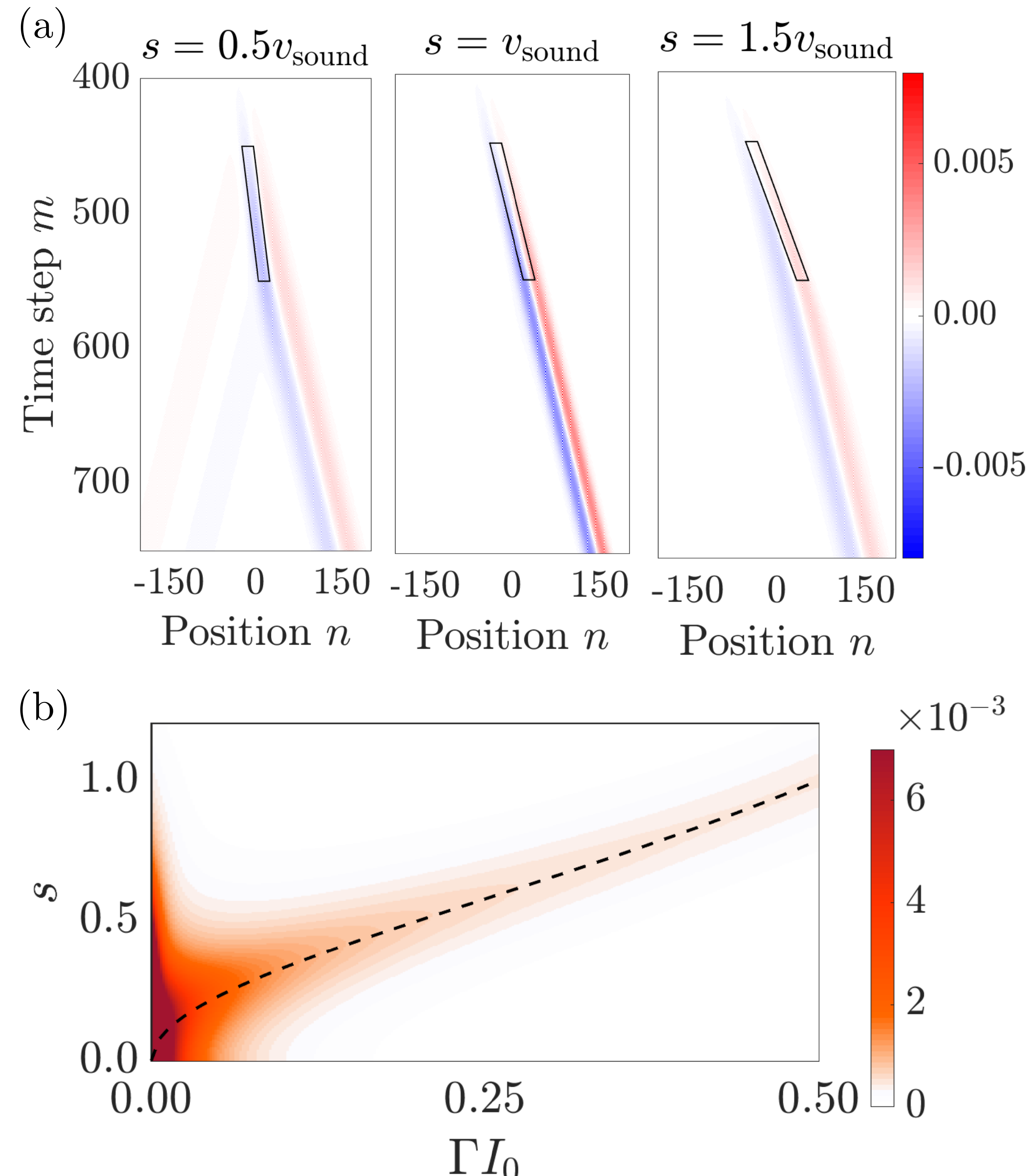}
	\caption{(a) Spatio-temporal colorplot of the numerical differential intensity (Eq.~\ref{eq:diff}) for a smoothly-varying, weak and wide defect, moving at speeds $s=\{0.5, 1, 1.5\} v_{\rm{sound}}$ across a uniform light field, with nonlinearity $\Gamma I_0 =0.25$, and with all other parameters as in Fig.~\ref{theory_sound}(a). The smooth defect profile (indicated by the black parallelogram) predominantly excites sound waves, with the strongest emission of radiation for $s= v_{\rm{sound}}$. (b) The total emission intensity (Eq.~\ref{eq:observable}), plotted as function of the defect velocity and nonlinearity for $m_{\rm{max}}=1000$ and $j=101$. Excitations are mostly suppressed when the defect moves slower than the speed of sound (Eq.~\ref{eq:sound}, \textit{dashed black line}). Note that little emission is also observed for defect speeds above the speed of sound, as then the defect is most resonant with higher-momenta excitations which remain relatively suppressed for wide defects [c.f. Eq.~\ref{eq:defectfourier}]. 	 }\label{linearmotionsimulated}	
\end{figure}

\subsubsection{Wide defect}
We then move to the case of a wide and smoothly-varying defect. 
The resulting spatio-temporal differential intensity is shown in Fig.~\ref{linearmotionsimulated}(a) for three different defect speeds; as the defect is very wide and slowly-varying (with the same parameters as in Fig.~\ref{theory_sound}(a)), we are in the regime where the defect predominantly excites sound waves. The strength of emission is then strongest when the defect moves at the speed of sound, such that these excitations are resonant. 
This is clearly visible in Fig.~\ref{linearmotionsimulated}(b), which shows how the total emission~\eqref{eq:observable} is indeed peaked around the speed of sound (dashed black line).
%
\begin{figure}[!]
	\includegraphics[width=0.95\linewidth]{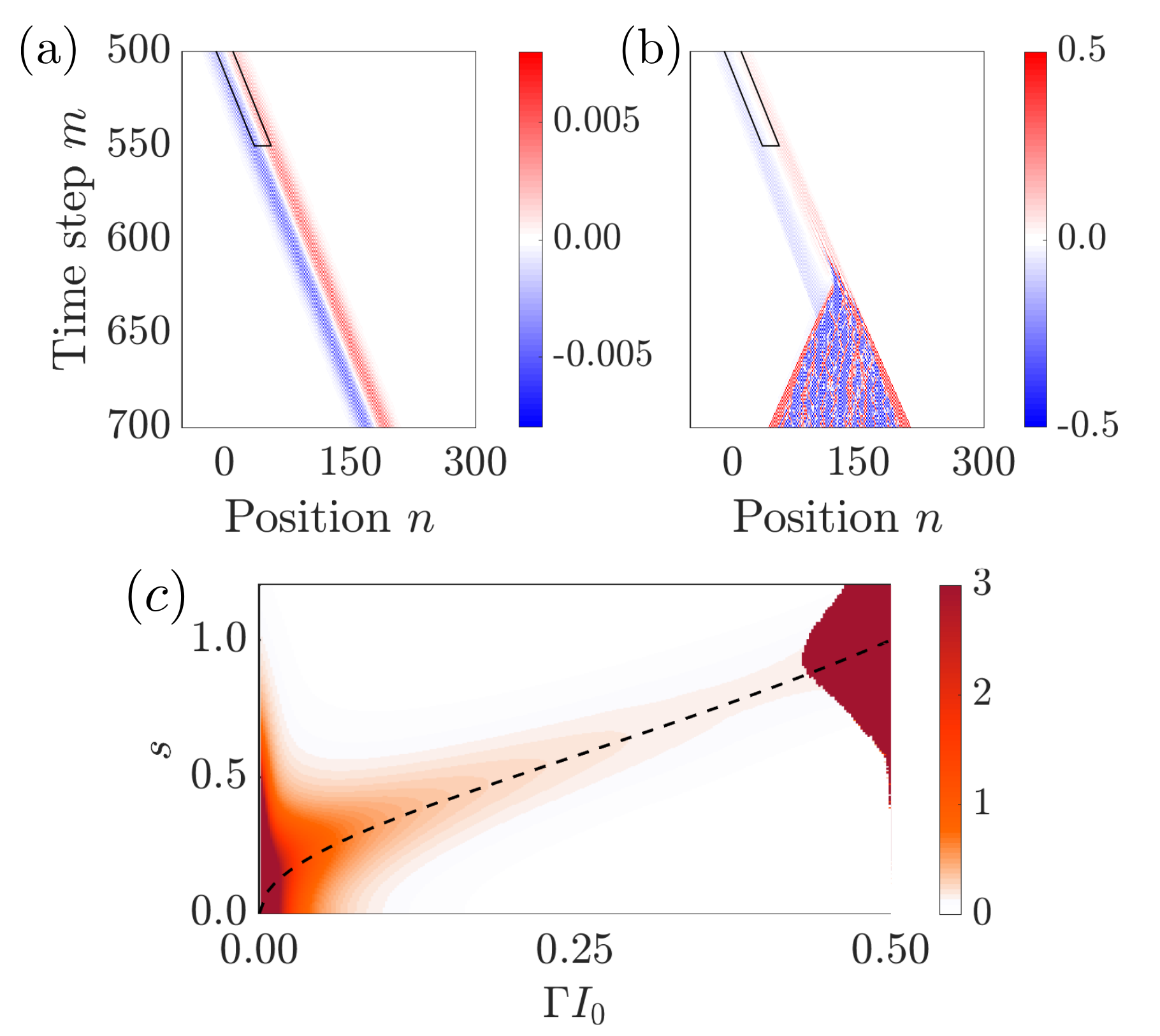}
	\caption{(a)\&(b) Spatio-temporal colorplot of the numerical differential intensity (Eq.~\ref{eq:diff}) for, respectively, a weak defect with amplitude $\varphi_0=0.01$, and a strong defect with amplitude $\varphi_0=0.2$. The defect (indicated by the black parallelogram) moves at a speed $s= v_{\rm{sound}}$ in a uniform light field, with nonlinearity $\Gamma I_0 =0.45$, and with all other parameters as in Fig.~\ref{theory_sound}(a). The strong defect in (b) triggers an instability, corresponding to a cascade of emitted excitations originating from the positive density bump that is present in the excitation pattern emitted by the defect. (c) The total emission intensity (\ref{eq:observable}) for the strong defect in (b), plotted as function of the defect velocity and nonlinearity for $m_{\rm{max}}=1000$ and $j=101$. The instability appears at large nonlinearities, spreading from the regime, $\Gamma I_0 >0.5$, where the light field is itself unstable. Note that the scale of the colorbar is chosen to highlight both stable and unstable regions; in fact, the emission intensity for the latter case is several orders of magnitude larger than for the former, and diverges further with increasing $m_{\rm{max}}$. }\label{instability}	
\end{figure}

Interestingly, by increasing the defect amplitude, we can also clearly observe the onset of the instability discussed in Sec.~\ref{sec:bog} for a defocusing nonlinearity. This is shown in Figure~\ref{instability} (a) and (b), where we compare the propagation of a weak and strong defect through a uniform light field at a high nonlinearity $\Gamma I_0 =0.45$. As can be seen, both defects emit sound waves; however, in the case of the strong defect, the system becomes unstable, leading to a cascade of emitted excitations. We note that the instability does not appear to be triggered by the defect itself, but instead by the positive density bump that is present in the density modulation emitted by the defect. This suggests that the instability arises as a result of the local light intensity, $I_{\rm{eff}}$, increasing sufficiently to enter the unstable regime $\Gamma I_{\rm{eff}}>0.5$, shown in Fig.~\ref{bogdispersion}. This is also consistent with Figure~\ref{instability} (c), where we plot the emission for the strong defect in (b) as a function of the defect speed and nonlinearity. Unlike for a weak defect [Fig.~\ref{linearmotionsimulated} (b)], we observe an unstable region of parameter space, which extends out from $\Gamma I_{\rm{0}}=0.5$, centered along the speed of sound and which pushes down to even lower nonlinearities as the defect amplitude is further increased.

\subsection{Superfluid-like response of expanding light fields} \label{sec:explandau}

As introduced in Section~\ref{sec:expan}, the initial light field in a realistic experiment is typically spatially-localised and then, unless  confining potentials are applied, freely expands over time. In this final section, we simulate a moving defect in these more realistic experimental conditions, to show that also here signatures of a superfluid-like behaviour and a critical velocity can be observed.

\begin{figure}[t]
	\includegraphics[width=0.9\linewidth]{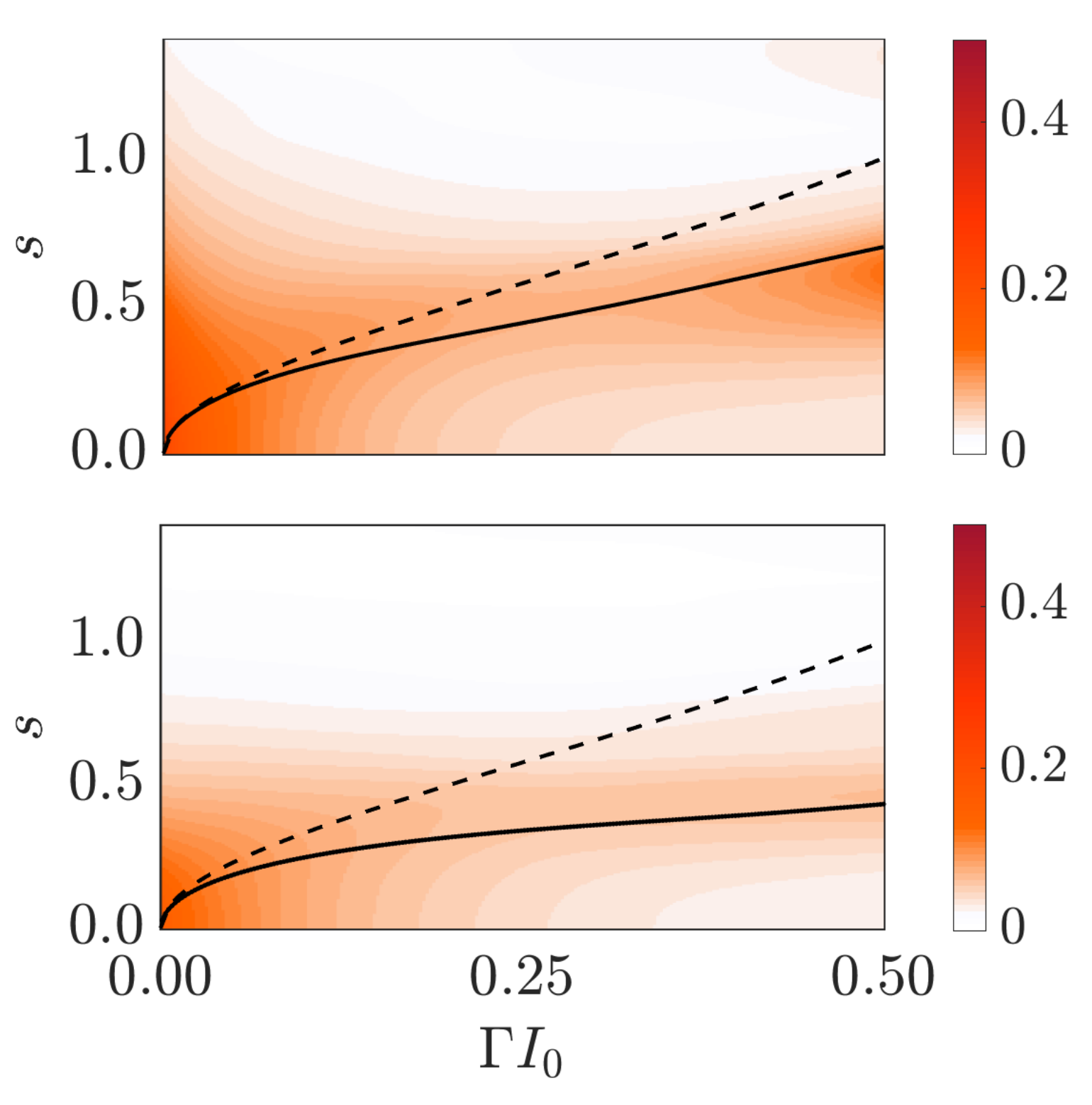}
	\caption{ The total emission intensity~\eqref{eq:observable} for a narrow and rapidly-varying defect, moved across a Gaussian light field with initial width $\sigma_G =10$ ({\it Top Panel}) and $\sigma_G =5$ ({\it Bottom Panel}). The defect parameters are $\sigma_n=1$, $\sigma_m=10$, $m_d=20$ and $\varphi_0=0.01$. The total emission intensity ~\eqref{eq:observable} has been evaluated with $m_{\rm{max}}=150$ and $j=51$. The analytical speed of sound from the initial intensity is shown as a dashed black line, and the speed of sound for the unperturbed intensity at the position of the defect is shown as a solid black line. As can be seen, the maximum of emission is in better agreement with the latter rather than the former, showing the impact of the light field expansion in reducing the critical speed for superfluidity.}\label{expanding}	
\end{figure}

In the two panels of Fig.~\ref{expanding}, we plot the total differential intensity~\eqref{eq:observable} for a realistic narrow and rapidly-varying defect, across a Gaussian light field with initial widths $\sigma_G =10$ and $\sigma_G =5$. The defect profile is chosen to have $\sigma_n=1$, $\sigma_m=10$; the rapid switch on and off of the defect leads to significant additional excitations, which smear out the threshold. In addition, the expansion of the light field also leads to a clear downwards shift  of the emission pattern towards lower speeds $s$ as compared to Fig.\ref{narrow}.

This can be attributed to two effects; firstly, the dropping intensity of the light field means that the effective nonlinearity at the time when the defect is applied is always lower than the initial nonlinearity at the center of the cloud. Here, we assess the importance of this effect by plotting the analytical speed of sound based on the initial intensity as a dashed black line, and the speed of sound for the unperturbed intensity at the central position of the defect as a solid black line; as expected, the maximum of emission is in much better agreement with the latter rather than the former, showing that the effective nonlinearity is indeed lower than would be expected from initial conditions. Secondly, the defect moves across the spatially inhomogeneous light field, and so also crosses regions of lower intensities. This leads to a sizable emission also below the expected threshold speed, as is indeed observed. Once all these complications are taken into account, the main qualitative features of the total emission intensity diagram of Fig.\ref{expanding} can be recognized and understood, confirming the important role of superfluid effects also in this expanding geometry.

While these conclusions are in qualitative agreement with the observations presented in the Supplemental Material of our previous experimental work Ref.~\onlinecite{experiment}, note that an alternative approach eventually turned out to be more effective in providing experimental evidence of a superfluid behaviour. As is presented in the main text of Ref.~\onlinecite{experiment}, expansion of the cloud could be blocked by applying an additional confining potential and the defect was then periodically moved through the cloud with a sinusoidal temporal dependence. This ensured that the light intensity did not drop significantly over time and that the defect motion was restricted to the central region of the cloud. This allowed for the experimental observation of a clear threshold in the deposited energy as a function of the defect speed, which was interpreted as evidence of the breakdown of superfluid behaviour above a certain critical speed. While this observation is in qualitative agreement with a generalized Landau criterion based on the instantaneous speed and the local density, a quantitative analysis of the experiment will require more sophisticated theoretical tools, in particular to account for the non-uniform motion of the defect and the consequent effect of acceleration on the emission. A study along these lines will be the topic of future work.

\section{Conclusions} \label{sec:conclusions}

In this paper, we have reported the development of a general theory of hydrodynamic effects in fluids of light in optical mesh lattices. This study had a twofold objective in mind.

On the one hand, our theory has allowed us to characterize effects, such as the behaviour of the speed of sound, the role of Umklapp processes in weakening superfluidity and the appearance of dynamical instabilities, which are qualitatively different from that found in typical superfluids of material or luminous particles in spatially continuous geometries, and are peculiar to the spatio-temporally periodic geometry of optical mesh lattices. 

On the other hand, we have 
performed detailed numerical simulations to show that these features can be indeed accurately measured in an optical mesh lattice, while taking into account experimental complications such as the overall expansion of the light field and realistic defect profiles. In particular, this study provides a solid conceptual framework in support of  the data analysis protocols applied in our recent experimental work~\cite{experiment}, and lays the foundation for a next generation of quantitative experimental measurements.
In particular, our work paves the way for further experimental exploration of more subtle superfluid hydrodynamic phenomena, going beyond our recent proof-of-principle experimental demonstration~\cite{experiment} and exploiting at full the novel and specific features of optical mesh lattices. For instance, our theoretical study has shown that new phenomena such as dynamical instabilities and effective critical velocities below the speed of sound may be observed in this platform.

As compared to other promising platforms for studies of superfluid hydrodynamics effects such as cold atoms~\cite{pitaevskii2016bose} or exciton-polariton fluids~\cite{RevModPhys.85.299}, optical mesh lattices have demonstrated direct real-time and site-resolved access to the fluid observables as well as an analogous spatio-temporal resolution in the design and application of external potentials with arbitrary shapes without the need for any cumbersome experimental apparatus. These considerations, together with the recent demonstration of lattices with non-trivial geometrical~\cite{wimmer2017experimental} or topological properties~\cite{bisianov,weidemann2020topological}, and of lattices with two~\cite{muniz} or even more spatial dimensions~\cite{RevModPhys.91.015006,ozawa2019topological}, suggest the promise of optical mesh lattices as an ideal candidate to investigate unprecedented regimes of superfluid hydrodynamics and of its interplay with the most subtle geometrical and topological features of complex lattices.

\begin{acknowledgments}
	This project was supported by German Research
	Foundation (DFG) in the framework of project PE 523/14-1, the Cooperative Research Center SFB 1375 NOA, and by the  International Research
	Training Group (IRTG) 2101.
	HMP is supported by the Royal Society via grants UF160112, RGF\textbackslash{}EA\textbackslash{}180121 and RGF\textbackslash{}R1\textbackslash{}180071. IC acknowledges financial support from the European Union FET-Open grant ``MIR-BOSE'' (n. 737017), from the H2020-FETFLAG-2018-2020 project "PhoQuS" (n.820392), from the Provincia Autonoma di Trento, from the Q@TN initiative, and from Google via the quantum NISQ award. 
\end{acknowledgments}

\end{document}